\documentclass[9pt,shortpaper,twoside,web]{ieeecolor}
\usepackage{generic}
\usepackage{cite}
\usepackage{amsmath,amssymb,amsfonts}
\usepackage{algorithmic}
\usepackage{graphicx}
\usepackage{textcomp}

 \usepackage{url} 
\usepackage{algorithmic}
\usepackage{algorithm}
\usepackage{array}
\usepackage{multirow}
\usepackage{booktabs}
\usepackage{nccmath}
\usepackage[caption=false,font=footnotesize]{subfig}

\def\BibTeX{{\rm B\kern-.05em{\sc i\kern-.025em b}\kern-.08em
    T\kern-.1667em\lower.7ex\hbox{E}\kern-.125emX}}
\begin{document}
\title{Differentially Private Perturbed Push-Sum Protocol and Its Application in Non-Convex Optimization}
\author{Yiming Zhou, Kaiping Xue, \IEEEmembership{Senior Member, IEEE}, Enhong Chen, \IEEEmembership{Fellow, IEEE}
\thanks{Y. Zhou is with the School of Artificial Intelligence and Data Science, University of Science and Technology of China, Hefei, Anhui, China.}
\thanks{K. Xue is with the School of Cyber Science and Technology, University of Science and Technology of China, Hefei, Anhui, China.}
\thanks{E. Chen is with the School of Computer Science and Technology, University of Science and Technology of China, Hefei, Anhui, China.}
\thanks{e-mail: zym2019@mail.ustc.edu.cn; \{kpxue, cheneh\}@ustc.edu.cn;}
\thanks{Corresponding author: E. Chen.}
}

\maketitle

\begin{abstract}
In decentralized networks, nodes cannot ensure that their shared information will be securely preserved by their neighbors, making privacy vulnerable to inference by curious nodes. Adding calibrated random noise before communication to satisfy differential privacy offers a proven defense; however, most existing methods are tailored to specific downstream tasks and lack a general, protocol-level privacy-preserving solution. To bridge this gap, we propose Differentially Private Perturbed Push-Sum (DPPS), a lightweight differential privacy protocol for decentralized communication. Since protocol-level differential privacy introduces the unique challenge of obtaining the sensitivity for each communication round, DPPS introduces a novel sensitivity estimation mechanism that requires each node to compute and broadcast only one scalar per round, enabling rigorous differential privacy guarantees. This design allows DPPS to serve as a plug-and-play, low-cost privacy-preserving solution for downstream applications built on it. To provide a concrete instantiation of DPPS and better balance the privacy–utility trade-off, we design PartPSP, a privacy-preserving decentralized algorithm for non-convex optimization that integrates a partial communication mechanism. By partitioning model parameters into local and shared components and applying DPPS only to the shared parameters, PartPSP reduces the dimensionality of consensus data, thereby lowering the magnitude of injected noise and improving optimization performance. We theoretically prove that PartPSP converges under non-convex objectives and, with partial communication, achieves better optimization performance under the same privacy budget. Experimental results validate the effectiveness of DPPS’s privacy-preserving and demonstrate that PartPSP outperforms existing privacy-preserving decentralized optimization algorithms. The code of experiments is available.\footnote{\url{https://github.com/WearTheClo/PartPSP}}
\end{abstract}


\section{Introduction}
\IEEEPARstart{T}{he} Perturbed Push-Sum protocol~\cite{OPS,PerturbedPS,AWPS} has emerged as a powerful decentralized communication and consensus mechanism in distributed optimization, particularly well-suited for settings where nodes interact over directed or time-varying networks. However, in decentralized networks where node origins and trustworthiness are often heterogeneous or unknown, each node cannot reliably assume that its neighbors will protect the information it shares. In response, the research community has explored various privacy-preserving techniques for decentralized systems, including state decomposition~\cite{StateDecom,StateDecomDP}, homomorphic encryption~\cite{PartHC1,PartHC2}, differential privacy~\cite{wang2017differential,DecLDP,katewa2019differential,xiong2020privacy,wang2023tailoring,DecGT-LDP}, and other insightful approaches~\cite{wang2023decentralized,wang2022quantization,wang2022privacy}, yielding significant advances in safeguarding sensitive data. Among these efforts, a notable line of works~\cite{PushSUMDP,NashPS,chen2023privacy,ADPVRSGP,gao2018privacy,wang2021privacy,PushStateDecom,zhang2022privacy} has adopted the Perturbed Push-Sum protocol as a communication backbone and integrated privacy mechanisms tailored to specific downstream tasks. While effective in their respective contexts, these insightful approaches typically couple the privacy mechanism tightly to the optimization objective or update rule, limiting the transferability of their theoretical findings and design principles to other downstream applications. 

To overcome those limitations, we advocate for embedding privacy directly into the communication protocol itself: a plug-and-play, protocol-level primitive that endows Perturbed Push-Sum with intrinsic, task-agnostic privacy guarantees. Differential privacy~\cite{DPpropose,DPDL} presents a natural foundation for this vision. The differential privacy provides a unified framework of strong formal privacy guarantees for mechanisms satisfying its definition, and has been widely adopted in distributed learning~\cite{wei2020federated,xue2023differentially,miao2024efficient,DPagainstMIA} to defend against inference attacks~\cite{MIAPropose,geiping2020inverting,MIALabelOnly,MIADiffusion} by honest-but-curious nodes. Its deployment is operationally simple, requiring only calibrated noise injection into outgoing messages based on sensitivity and a privacy budget, without cryptographic overhead or complex coordination. It is precisely this modular and lightweight deployment that makes differential privacy a highly attractive candidate for building protocol-level privacy into the Perturbed Push-Sum protocol. Nevertheless, despite its clear promise, realizing this vision remains hindered by significant theoretical and algorithmic challenges.

The primary challenge in endowing the Perturbed Push-Sum protocol with differential privacy lies in accurately estimating its sensitivity during execution. The sensitivity indicates the maximum difference between the output parameters and determines the magnitude of the added noise. A higher sensitivity value represents greater diversity in the communicated parameters, necessitating the addition of larger noise to obscure the differences. For instance, in differentially private optimization algorithms, the sensitivity is typically tied to the gradient clipping norm~\cite{DPDL,wei2020federated}, as the goal is to protect individual data samples in the local datasets. In contrast, for the Perturbed Push-Sum protocol operating in decentralized networks, sensitivity must be derived from the worst-case deviation distance between each node’s communicated parameters, since the objective is to provide privacy for the information each node sends out. More specifically, satisfying differential privacy requires each node to know the maximum norm distance between its outgoing parameters and those of all other nodes to properly add privacy-preserving noise. However, this requirement is inherently difficult to fulfill in a decentralized network, where nodes lack global knowledge of the information sent by others.

The second key challenge lies in demonstrating the practical utility of the differentially private Perturbed Push-Sum protocol in downstream tasks. The primary motivation for embedding differential privacy directly into the protocol is to provide a plug-and-play privacy guarantee, enabling algorithms built upon it to inherit provable privacy protection without redesigning its core method. This raises a critical question: can such a privacy-preserving protocol still support efficient and stable execution of demanding downstream tasks? Among these, non-convex optimization stands out as a particularly important application, given its central role in machine learning. However, existing literature~\cite{wang2017differential,wei2020federated,xiong2020privacy,xue2023differentially} consistently shows that the noise required for differential privacy inevitably degrades convergence performance, forcing a fundamental trade-off between privacy protection and optimization performance. Consequently, a major open problem is how to design non-convex optimization algorithms based on the differentially private Perturbed Push-Sum protocol that mitigate the adverse impact of privacy-preserving noise, thereby achieving the best possible convergence guarantees under a given privacy budget. Addressing this challenge is essential to realizing the full potential of protocol-level privacy in real-world decentralized optimization systems.

To deal with the above challenges, we propose the \textbf{D}ifferentially \textbf{P}rivate Perturbed \textbf{P}ush-\textbf{S}um (\textit{DPPS}) protocol and build upon it a decentralized optimization algorithm, called \textbf{Part}ial Communication \textbf{P}ush-\textbf{S}um SGD with Differential \textbf{P}rivacy (\textit{PartPSP}). First, to overcome the inaccessibility of global sensitivity in decentralized networks, we introduce a lightweight sensitivity estimation method for the DPPS protocol. Each node $i$ in the decentralized network calculates its sensitivity $S_i\in \mathbb{R} $ relative to the network based on its information. We theoretically prove that using the maximum $S_i$ across the network as the sensitivity for the current communication ensures that the DPPS protocol satisfies differential privacy. This design achieves provable privacy with negligible additional overhead, making DPPS highly practical for real-world deployment. Second, our analysis of the DPPS protocol shows that its sensitivity grows with the dimension of the shared parameters, increasing the required noise magnitude and hurting downstream performance. To address this, PartPSP adopts a partial communication mechanism~\cite{collins2021exploiting,pillutla2022federated,DFedPGP,chen2024fedbone} that splits the model into local and shared parameters, reducing the dimension of shared parameters and thus the sensitivity for differential privacy. A convergence analysis for non-convex objectives shows that this approach maintains optimization guarantees and improves the privacy–utility trade-off: under the same privacy budget, PartPSP achieves better convergence than full-communication baselines. We verify these results experimentally on deep learning tasks. The contributions of this paper are summarized as follows.

\begin{itemize}
    \item We propose the DPPS protocol, which achieves differential privacy via the Laplace mechanism and thereby provides protocol-level privacy guarantees for all nodes in the network. To address the challenge of sensitivity estimation, we propose a lightweight scheme in which each node computes a scalar and broadcasts it to the network, then the protocol adopts the maximum value as the sensitivity for this communication round. We formally prove that, with this low-cost sensitivity estimation, DPPS satisfies differential privacy, thus offering task-agnostic and provable privacy protection to downstream applications built upon it.
    \item We propose PartPSP, an exemplary downstream instantiation of the DPPS protocol for decentralized non-convex optimization. Building upon DPPS, PartPSP introduces a partial communication mechanism that partitions model parameters into local and shared components, applying DPPS only to the shared ones for private consensus. This reduces the dimensionality of the noised parameters, thereby lowering the required noise magnitude and improving optimization performance. Our convergence analysis under non-convex objectives not only establishes PartPSP’s convergence but also demonstrates that, under the same privacy budget, it achieves better optimization guarantees.
    \item We evaluate our approach on real-world datasets using widely adopted deep learning tasks. We validate DPPS’s lightweight sensitivity estimation in practice, thereby empirically verifying its privacy guarantees under realistic conditions, and further investigate how theoretically identified factors, such as network topology and partial communication, affect sensitivity during training. We further demonstrate that PartPSP, equipped with partial communication, achieves significantly better optimization performance than existing privacy-preserving decentralized algorithms under the same privacy budget, effectively alleviating the utility degradation caused by differential privacy.
\end{itemize}

We organize the remainder of this paper as follows. Section II formalizes the problem setting under study. Section III presents our main contributions: the DPPS protocol, its lightweight sensitivity estimation method, and PartPSP. Section IV provides the theoretical guarantees on privacy and convergence. Section V empirically validates our theoretical findings through extensive experiments. Finally, Section VI concludes the paper.
\begin{table}[t]
\centering
\caption{Definitions of mathematical symbols}
\begin{tabular}{cc}
    \toprule
    \textbf{Symbol} & \textbf{Description} \\
    \midrule
    $N$ & the number of nodes in the decentralized network \\
    $[N]$ & $\{1, 2, ..., N\}$, is the set of nodes \\
    $\mathcal{G} ^{(t)}$ & the network graph at time $t$ \\
    $\textbf{W}^{(t)}$ & the weight matrix generated by graph $\mathcal{G} ^{(t)}$\\
    $\textbf{s}_i^{(t)}$ & the vector of shared parameters of node $i$ at time $t$\\
    $\textbf{l}_i^{(t)}$ & the vector of local parameters of node $i$ at time $t$\\
    $\textbf{n}_i^{(t)}$ & the noise vector with the same dimension as $\textbf{s}_i^{(t)}$\\
    $d_s$ & the number of dimension of $\textbf{s}$\\
    $\bar{\textbf{s}}$ & the vector of average shared parameters of the network\\
    $\textbf{L}$ & the set of all $\textbf{l}$ in the network\\
    $b$ & the hyperparameter of privacy budget\\
    $Lap(0, \frac{S}{b})$ & the Laplace distribution with parameters 0 and $\frac{S}{b}$\\
    $\left \| \textbf{a}  \right \| _1$ & the L1 norm of vector \textbf{a} \\
    $\left \| \textbf{a}  \right \| _2$ & the L2 norm of vector \textbf{a} \\
    \bottomrule
\end{tabular}
\end{table}

\section{Preliminaries}
In this section, we introduce the knowledge of the decentralized network properties, the concept of differential privacy, and the optimization of distributed learning. Table \uppercase\expandafter{\romannumeral 1} lists the important mathematical symbols used in this paper.

\subsection{Network Settings}
For a decentralized network with node scale $N$, its network topology can be described by a sequence of directed graphs $\left\{\mathcal{G} ^{(t)}\right\}=\left\{\left[N\right],E^{(t)}\right\} $, where $\left[N\right]=\left\{ 1,2,...,N\right\}$ is the set of nodes and $E^{(t)}\subseteq\left[N\right]\times\left[N\right]$ is the set of communication links. A link $(i,j)\in E^{(t)}$ indicates that node $i$ has a reliable one-way communication channel for sending messages to node $j$ during time $t$. We introduce the following connectivity assumption to the network topology.

\noindent
{\bf Assumption 1}.{ \it
By convention, each node in the graph sequence $\left\{\mathcal{G} ^{(t)}\right\}$ has a self-loop. There exist finite positive integers $B$ and $\Lambda$, such that the aggregate graph $\bigcup_{i=t}^{t+B-1}\mathcal{G} ^{(i)}$ is strongly connected and each aggregate graph has a diameter at most $\Lambda$ for every $t\ge 0$.
}

\noindent
Assumption 1 is brought from the Perturbed Push-SUM protocol~\cite{PerturbedPS,AWPS} and limits the network's connectivity.

Each graph $\mathcal{G} ^{(t)}$ would generate a weight matrix $\textbf{W}^{(t)}=\left [ w_{i,j}^{(t)} \right ] _{N\times N}$ by the following definitions.

\noindent
{\bf Definition 1}.{ \it
For every $t\ge 0$, each matrix $\textbf{W}^{(t)}=\left [ w_{i,j}^{(t)} \right ] _{N\times N}$ is a doubly stochastic matrix, and $w^{(t)}_{i,j} > 0$ if and only if $\left ( j,i \right ) \in E^{(t)}$.
}

\noindent
Notably, unlike the Perturbed Push-Sum protocol~\cite{PerturbedPS,AWPS} that only requires the weight matrices to be column-stochastic, the matrices considered in this work must be doubly stochastic to ensure the validity of our subsequent sensitivity estimation method.

\subsection{Security Model}

We employ differential privacy to provide privacy protection for each node through protocol-level design. Differential privacy is a property of the database. The database is a mapping that maps an input query to the corresponding answer set. At each communication round of the protocol, every node obtains the shared parameters $\textbf{s}$ of its neighbors by querying them. Modeling the network as a database, we have the following definitions.

\noindent
{\bf Definition 2}.{ \it
(Query). A query $q$ is a subset of the node set $\left [ N \right ] $ of the network. Denote by $Q$ the set of all possible queries.
}

\noindent
{\bf Definition 3}. {\it 
(Mapping). The mapping $m: \mathbb{R}^{|q|} \rightarrow \mathbb{R}^{|q| \times d_s}$ maps an input query $q$ to the corresponding nodes' shared parameters, defined as:

\begin{equation}\label{mapm}
m(q) = \left[ \textbf{s}^{(t+\frac{1}{2})} \right]_q,
\end{equation}

\noindent
where $\left[ \textbf{s}^{(t+\frac{1}{2})} \right]_q$ is a $\mathbb{R}^{|q| \times d_s}$ matrix that consists of the shared parameters $\textbf{s}_i^{(t+\frac{1}{2})}\in \mathbb{R}^{d_s}$ of all nodes $i \in q$.
}

\noindent
Next, we provide the following mathematical definition of differential privacy \cite{DPpropose} for a mapping.

\noindent
{\bf Definition 4}.{ \it
($\varepsilon$-Differential Privacy). A mapping M is $\varepsilon$-differential privacy if for all queries $q, q' \in Q$ which differ in only one entry, for all possible answer set $A$, we have:
\begin{equation}\label{DP}
\frac{Pr\left [ M(q)=A \right ] }{Pr\left [ M(q')=A \right ] } \le e^\varepsilon.
\end{equation}
}

\noindent
To transform the mapping $ m $ into a differentially private mapping $ M $, we introduce the following conceptions \cite{DPpropose}.

\noindent
{\bf Definition 5}.{ \it
($L_1$ Sensitivity). The $L_1$ sensitivity of a mapping $m: \mathbb{R}^{\left | q \right |}  \rightarrow \mathbb{R}^{\left | q \right |\times d_s}$ is the smallest number $S_m$ such that for all queries $q, q' \in Q$ which differ in only one entry,
\begin{equation}\label{Sensitivity}
\left \| m(q)-m(q') \right \|_1\le S_m.
\end{equation}
}

\noindent
{\bf Lemma 1}.{ \it
(Laplace Mechanism). For the given mapping $m: \mathbb{R}^{\left | q \right |}  \rightarrow \mathbb{R}^{\left | q \right |\times d_s}$ with $L_1$ sensitivity $S_m$, for any $\varepsilon > 0 $, the mapping M defined as follows ensures $\varepsilon$-differential privacy:
\begin{equation}\label{LaplaceM}
M\left ( q \right ) =m\left ( q \right ) + L_{noise},
\end{equation}
where $L_{noise}\in \mathbb{R}^{\left | q \right |\times d_s}$ consists of elements independently and identically distributed (i.i.d.) according to the Laplace distribution $Lap(0, \frac{S_m}{\varepsilon})$.
}

We consider a threat model in which an honest-but-curious node in the decentralized network makes only a limited number of attempts to infer private information of other nodes by observing their shared parameters through the protocol. Such inference attacks are generally not mitigated by standard cryptographic measures, as they exploit the content of legitimate communications rather than intercepting them. Enforcing differential privacy on the communication protocol provides a provable defense against this form of privacy leakage.

\subsection{Optimization Problem}
We consider an optimization problem equipped with a partial communication mechanism. Under this mechanism, each node’s model parameters are partitioned into two parts. One part is kept local and never communicated over the network; we refer to it as the local parameters $\textbf{l}$. The other part is allowed to be shared across the network to enable collaborative learning and is perturbed with a noise vector to ensure differential privacy; we refer to this part as the shared parameters $\textbf{s}$. We aim to solve the following problem.

\noindent
{\bf Problem 1}. { \it
The algorithm for a network with $N$ nodes is designed to solve the following minimization problem.
\begin{eqnarray*}\label{Problem}
&& \min_{\textbf{s}, \textbf{L}} \left \{ F\left (\textbf{s},\textbf{L}\right ):= \frac{1}{N}\sum_{i=1}^{N}F_{i}\left (\textbf{s},\textbf{l}_i\right ) \right \},\\
&&~~~~~~ where\ F_{i}\left (\textbf{s},\textbf{l}_i\right ):= \mathbb{E}_{\xi_{i}\sim D_{i}} \left [ F_{i}\left ( \textbf{s},\textbf{l}_i, \xi_{i}\right ) \right ]. 
\end{eqnarray*}
}

\noindent
For Problem 1, the $\textbf{L}$ is a set of all local parameters, $F_{i}\left (\textbf{s},\textbf{l}_i\right )$ is the expected local objective function, and $\xi_{i}$ is a data sampling according to the local dataset $D_{i}$. Our algorithm aims to find a unified shared parameters $\textbf{s}$ for the network and a personalized local parameters $\textbf{l}_i$ for each node to minimize the average expected local loss value. We define the following average parameters for the optimization.

\noindent
{\bf Definition 6}.{ \it
The average shared parameters in the network is defined as $\bar{\textbf{s}}=\frac{1}{N}\sum_{i=1}^{N}\textbf{s}_{i}$.
}

Our algorithm is based on the stochastic gradient descent to optimize Problem 1. We define the following stochastic gradient terms to represent different partial derivatives with respect to $\textbf{s}$ and $\textbf{l}$, respectively.

\noindent
{\bf Definition 7}.{ \it
We define the following stochastic partial derivatives for the parameters $\textbf{s}$ and $\textbf{l}$, respectively.
\begin{equation}
\textbf{g}_{i,l}^{(t)}:=\nabla _l F_i \left ( \textbf{y}_{i}^{(t)},\textbf{l}_{i}^{(t)};\xi ^{(t)}_{i} \right ),
\end{equation}
\begin{equation}
\textbf{g}_{i,s}^{(t)}:=\nabla _s^c F_i \left ( \textbf{y}_{i}^{(t)},\textbf{l}_{i}^{(t+1)};\xi ^{(t)}_{i} \right ),
\end{equation}
where $\xi_{i}$ is a stochastic sampling from the local dataset $D_{i}$.
}

\noindent
We make two adaptations to $\textbf{g}_{i,s}^{(t)}$. First, due to the requirements~\cite{SGPConvex,SGP} of the Perturbed Push-SUM protocol, we replace the shared parameters $\textbf{s}$ with their corrected counterparts $\textbf{y}$. These two parameters differ only by a constant multiple. Second, we apply the widely used gradient clipping technique \cite{DPDL} to limit the L1 norm of $\textbf{g}_{i,s}^{(t)}$, which will be explained later. In summary, $\nabla _s^c$ represents the clipped partial derivative of the parameters at the position of $\textbf{s}$.

For the objective function $F$, we make the following smoothness assumption.

\noindent
{\bf Assumption 2}.{ \it
(Smoothness). For every node $i\in[N]$, its local function $F_i$ is continuously differentiable. There exist constants $L_l$, $L_s$, $L_{ls}$, and $L_{sc}$ such that for every node $i\in[N]$:
\begin{itemize}
\item[(a)] $\left \|  \nabla _l F_i\left ( \textbf{s}, \textbf{l}_1  \right )-\nabla _l F_i\left ( \textbf{s}, \textbf{l}_2  \right )  \right \| _2\le L_l\left \| \textbf{l}_1-\textbf{l}_2 \right \| _2,$
\item[(b)] $\left \|  \nabla _l F_i\left ( \textbf{s}_1, \textbf{l}  \right )-\nabla _l F_i\left ( \textbf{s}_2, \textbf{l}  \right )  \right \| _2\le L_{ls}\left \| \textbf{s}_1-\textbf{s}_2 \right \| _2,$
\item[(c)] $\left \|  \nabla _s F_i\left ( \textbf{s}_1, \textbf{l}  \right )-\nabla _s F_i\left ( \textbf{s}_2, \textbf{l}  \right )  \right \| _2\le L_{s}\left \| \textbf{s}_1-\textbf{s}_2 \right \| _2,$
\item[(d)] $\left \|  \nabla _s F_i\left ( \textbf{s}_1, \textbf{l}  \right )-\nabla _s^c F_i\left ( \textbf{s}_2, \textbf{l}  \right )  \right \| _2\le L_{sc}\left \| \textbf{s}_1-\textbf{s}_2 \right \| _2.$
\end{itemize}
}

\noindent
Regarding the distribution of local datasets, we assume the following.

\noindent
{\bf Assumption 3}.{ \it
(Bounded Variance). There exist $\sigma_{s}>0$, $\sigma_{l}>0$ and $\sigma_{g}>0$ such that for all $\textbf{s}, \textbf{l}$ and $i\in [N]$, it follows:
\begin{itemize}
\item[(a)] $\mathbb{E}\left [ \left \| \nabla _s^c F_i \left ( \textbf{s}_i,\textbf{l}_i; \xi_{i}\right )- \nabla _s^c F_i \left ( \textbf{s}_i,\textbf{l}_i\right )\right \|_2^2 \right ]\le \sigma_{s}^2 $,
\item[(b)] $\mathbb{E}\left [ \left \| \nabla _l F_i \left ( \textbf{s}_i,\textbf{l}_i; \xi_{i}\right )- \nabla _l F_i \left ( \textbf{s}_i,\textbf{l}_i\right )\right \|_2^2 \right ]\le \sigma_{l}^2 $,
\item[(c)] $\mathbb{E}\left [ \left \| \nabla _s^c F_i \left ( \textbf{s}_i,\textbf{l}_i\right ) - \nabla _s^c F \left ( \textbf{s}_i,\textbf{L}\right )\right \|_2^2 \right ]\le \sigma_{g}^2 $.
\end{itemize}
}

\noindent
The above two assumptions are variants of conditions widely used in non-convex optimization~\cite{SGP,ADPVRSGP} and will be employed in the theoretical convergence analysis of PartPSP.
\section{Methodology}
\subsection{DPPS Protocol} 
\begin{algorithm}
\caption{Differentially Private Perturbed Push-Sum Protocol}
\textbf{Input}: Number of iterations $T$, noise rate $\gamma_n$, privacy budget hyperparameter $b$. Each node $i$ initializes its shared parameters $\textbf{s}_{i}^{(0)}$, corrected parameters $\textbf{y}_{i}^{(0)}=\textbf{s}_{i}^{(0)}$, normalizing scalar $a^{(0)}_{i}=1$.\\
\textbf{Output}: $\bar{\textbf{s}}^{(T)}$.
\begin{algorithmic}[1] 
\FOR{$t=\{0, 1, ..., T-1\}$}
\FOR{each node $i\in [N]$ in parallel}

\STATE Update $\textbf{s}$ with  the perturbation $\varepsilon ^{(t)}_{i}$:
\begin{equation}\label{Perturbed}
\textbf{s}_{i}^{(t+\frac{1}{2} )} = \textbf{s}_{i}^{(t)} + \varepsilon ^{(t)}_{i} ;
\end{equation}

\STATE Compute its $S_i^{(t)}$ using Eq.~\eqref{iterSen}, broadcasts $S_i^{(t)}$ to the network, and obtains the sensitivity $S^{(t)} = \max_{i \in [N]} S_i^{(t)}$. 

\STATE Sample the noise vector $\textbf{n}_i^{(t)} \in \mathbb{R}^{d_s}$ from the Laplace distribution $Lap(0, \frac{S^{(t)}}{b})$ and adds it to $\textbf{s}$ as follows:
\begin{equation}\label{AddNoise}
\textbf{s}_{i, noise}^{(t+\frac{1}{2})} = \textbf{s}_{i}^{(t+\frac{1}{2})}+\gamma_n\textbf{n}_{i}^{(t)};
\end{equation}

\STATE Send $\textbf{s}_{i, noise}^{(t+\frac{1}{2})}$, $a^{(t)}_{i}$, and $w^{(t)}_{j,i}$ to each neighbor $j$ and receives the corresponding messages.

\STATE Average the received messages via:
\begin{equation}\label{Consensus}
\left\{
\begin{array}{lr}
a_{i}^{(t+1)}=\sum_{(j,i)\in E^{(t)}} w_{i,j}^{(t)}a_{j}^{(t)};  \\
\textbf{s}_{i}^{(t+1)}=\sum_{(j,i)\in E^{(t)}} w_{i,j}^{(t)}\textbf{s}_{j, noise}^{(t+\frac{1}{2})};
\end{array}
\right.
\end{equation}

\STATE Update the corrected parameters $\textbf{y}$ via:
\begin{equation}\label{corrected}
\textbf{y}_{i}^{(t+1)}=\textbf{s}_{i}^{(t+1)}/a_{i}^{(t+1)};
\end{equation}

\ENDFOR
\ENDFOR
\end{algorithmic}
\end{algorithm}

The DPPS protocol, as presented in Algorithm 1, provides a general framework for private decentralized consensus. At each iteration $t$, each node $i\in[N]$ first adds a local perturbation $\varepsilon_i ^{(t)}$ to its shared parameters $\textbf{s}_i^{(t)}$ (Line 3), then computes its local sensitivity $S_i^{(t)}\in \mathbb{R}$ based on current information (Line 4). By broadcasting their local sensitivities and taking the maximum $S^{(t)} = \max_{i \in [N]} S_i^{(t)}$ across the network, all nodes obtain a common sensitivity value $S^{(t)}$, which is used to generate Laplace noise for differential privacy (Line 5). The noised parameters $\textbf{s}_{i, noise}^{(t+\frac{1}{2})}$ are then exchanged over the network and aggregated by each node (Lines 6–7). Finally, a normalized scalar is used to correct the parameters and enable consensus (Line 8).

Compared to the Perturbed Push-Sum protocol, DPPS introduces only two additional steps: sensitivity estimation (Line 4) and noise injection (Line 5). When the dimension of the local perturbation matches that of the shared parameters, each node computes its local sensitivity in time $O(d_s)$ using only locally stored information; details of the computation are provided in the next subsection. The resulting scalar is then broadcast across the network, incurring a communication cost of $O(N)$, to obtain the sensitivity $S^{(t)}$ required for differential privacy. In Line 5, generating and adding Laplace noise incurs an additional computational cost of $O(d_s)$. Thus, relative to Perturbed Push-Sum, DPPS adds only $O(d_s)$ computation and $O(N)$ communication overhead per iteration—overhead that is negligible compared to the cost of computing the perturbation itself and communicating the $d_s$-dimensional shared parameters.

\subsection{Sensitivity Estimation}

For the DPPS protocol, to ensure that the shared parameters $\textbf{s}_i^{(t+\frac{1}{2})}$ sent by node $i$ at time $t$ are privacy-protected, we need to add noise to them to satisfy differential privacy. According to Definition 3, the mapping $ m $ formally describes the process in which each node sends its shared parameters to a querying entity without any processing; in other words, $ m $ corresponds to the noiseless version of the DPPS protocol at iteration $t$, where shared parameters are transmitted without noise. To construct a differentially private mapping $M$ for node $i$ from mapping $m$, where $M$ is the complete DPPS protocol described in Algorithm 1, Lemma 1 requires us to compute the sensitivity of $m$. By Definition 5, the sensitivity of the mapping $m$ at iteration $t$ is straightforward: $\max_{i,j \in [N]} \left\| \textbf{s}_i^{(t+\frac{1}{2})} - \textbf{s}_j^{(t+\frac{1}{2})} \right\|_1$. The following Lemma 2 provides an upper bound on this sensitivity, which serves as the basis for noise generation in the DPPS protocol.


\noindent
{\bf Lemma 2}.{ \it
(Sensitivity Estimation). Suppose that Assumption 1 holds and that the weight matrices $\mathbf{W}^{(t)}$ satisfy Definition 1. Then, when the noiseless version of the DPPS protocol at iteration $t > 0$ is identified as the mapping $m$, there exists an upper bound for $m$, given by:
\begin{equation*}\label{SBound}
\max_{i,j \in [N]} \left\|\textbf{s}^{(t+\frac{1}{2})}_i - \textbf{s}^{(t+\frac{1}{2})}_j \right\|_1 \leq \max_{i\in [N]}\left \{ S_i^{(t)} \right \} 
\end{equation*}
where $S_i^{(t)}$ is defined as:
\begin{equation}\label{SensPerIter}
\begin{aligned}
S_i^{(t)} = & 2C'\lambda^{t}\left\|\textbf{s}_{i}^{(0)} \right\|_1 + 2 C'\sum_{k=0}^{t}\lambda^{t-k}\left\|\varepsilon ^{(k)}_{i} \right\|_1 \\
& + 2\gamma_n C'\sum_{k=0}^{t-1}\lambda^{t-k}\left\|\textbf{n}_{i}^{(k)} \right\|_1,
\end{aligned}
\end{equation}
and where $C' > 0$ and $\lambda \in (0,1)$ are constants.
}

\begin{proof}
For iteration $t>0$, consider the $L_1$ norm distance between node $i$ and any node $j\in[N]$, we have:

\begin{equation*}\label{lemma21eq1}
\begin{aligned}
&&\left \|\textbf{s}^{(t+\frac{1}{2})}_i- \textbf{s}^{(t+\frac{1}{2})}_j \right \|_1 & \le \left \|\textbf{s}^{(t+\frac{1}{2})}_i- \textbf{y}^{(t+1)}_i+\textbf{y}^{(t+1)}_i- \bar{\textbf{s} }^{(t)}\right \|_1\\
&&&\quad +\left \|\textbf{s}^{(t+\frac{1}{2})}_j- \textbf{y}^{(t+1)}_j+\textbf{y}^{(t+1)}_j- \bar{\textbf{s} }^{(t)}\right \|_1.
\end{aligned}
\end{equation*}

\noindent
Considering the norm of node $i$ is larger, so that
\begin{equation}\label{lemma21eq2}
\begin{aligned}
&&&\left \|\textbf{s}^{(t+\frac{1}{2})}_i- \textbf{s}^{(t+\frac{1}{2})}_j \right \|_1\\
&&&\le2\left \|\textbf{s}^{(t+\frac{1}{2})}_i- \textbf{y}^{(t+1)}_i+\textbf{y}^{(t+1)}_i- \bar{\textbf{s} }^{(t)}\right \|_1\\
&&&\le2\underbrace{\left \|\textbf{s}^{(t+\frac{1}{2})}_i- \textbf{y}^{(t+1)}_i\right \|_1}_{(a)} + 2\left \| \textbf{y}^{(t+1)}_i- \bar{\textbf{s} }^{(t)}\right \|_1.
\end{aligned}
\end{equation}

\noindent
For (a) in Eq.~\eqref{lemma21eq2}, consider $ \textbf{y}_{i}^{(t+1)} $,
\begin{equation}\label{lemma21eq3}
\begin{aligned} 
&&\left ( \textbf{y}_{i}^{(t+1)} \right )^T & =\left (\frac{\sum_{(j,i)\in E^{(t)}} w_{i,j}^{(t)}\textbf{s}_{j,noise}^{(t+\frac{1}{2})}}{\sum_{(j,i)\in E^{(t)}} w_{i,j}^{(t)}a_{j}^{(t)}}\right )^T\\
&&&\overset{(a)}{= } \left (\frac{\sum_{(j,i)\in E^{(t)}} w_{i,j}^{(t)}\textbf{s}_{j}^{(t+\frac{1}{2})}}{\sum_{(j,i)\in E^{(t)}} w_{i,j}^{(t)}a_{j}^{(t)}}\right )^T\\
&&& =\frac{\left [ \textbf{W}^{(t)} \right ]_i \left [ \textbf{s}_{1}^{(t+\frac{1}{2})} ,  \dotsm, \textbf{s}_{N}^{(t+\frac{1}{2})}\right ]^{T} }{\left [ \textbf{W}^{(t)} \right ]_i \textbf{a}^{(t)}},
\end{aligned}
\end{equation}
where (a) in Eq.~\eqref{lemma21eq3} holds because the mapping $m$ outputs $\textbf{s}_{i}^{(t+\frac{1}{2})}$ without adding noise; thus, for $m$, we have $\textbf{s}_{i}^{(t+\frac{1}{2})} = \textbf{s}_{i,noise}^{(t+\frac{1}{2})}$, in contrast to the DPPS protocol. Here, $\textbf{a}^{(t)}$ denotes the vector of normalizing scalars $a_i^{(t)}$, and $\textbf{W}^{(t)}$ is generated according to Definition~1, with $\left [ \textbf{W}^{(t)} \right ]_i$ denoting its $i$-th row.

\noindent
By rearranging Eq.~\eqref{lemma21eq3}, we obtain:
\begin{equation}\label{lemma21eq4}
\textbf{a}^{(t)}\left ( \textbf{y}_{i}^{(t+1)} \right )^T=\left [ \textbf{s}_{1}^{(t+\frac{1}{2})} , \dots, \textbf{s}_{N}^{(t+\frac{1}{2})}\right ]^{T}.
\end{equation}

\noindent
Then, we have
\begin{equation}\label{lemma21eq5}
\textbf{s}_{i}^{(t+\frac{1}{2})} = a_{i}^{(t)}\textbf{y}_{i}^{(t+1)}.
\end{equation}

\noindent
For $a_{i}^{(t)}$, consider $\textbf{a}^{(t)}$,
\begin{equation}\label{lemma21eq6}
\textbf{a}^{(t)}=\textbf{W}^{(t)}\textbf{W}^{(t-1)} \dotsm\textbf{W}^{(0)}\textbf{a}^{(0)} \stackrel{(a)}{=}\textbf{1},
\end{equation}
where (a) follows from the fact that the matrices $\mathbf{W}^{(t)}$ are doubly stochastic, as stated in Definition~1.

\noindent
Then, for (a) in Eq. (\ref{lemma21eq2}), we have
\begin{equation}\label{lemma21eq7}
\left \|\textbf{s}^{(t+\frac{1}{2})}_i- \textbf{y}^{(t+1)}_i\right \|_1=\left \|\textbf{y}^{(t+1)}_i- \textbf{y}^{(t+1)}_i\right \|_1=0.
\end{equation}

\noindent
Substituting Eq.\eqref{lemma21eq7} back into Eq.\eqref{lemma21eq2} and considering the other case where the norm of $j$ is larger, we have:
\begin{equation}\label{lemma21eq_tem}
\max_{i,j \in [N]} \left \|\textbf{s}^{(t+\frac{1}{2})}_i- \textbf{s}^{(t+\frac{1}{2})}_j \right \|_1\le \max_{i\in[N]}\left \{2\left \| \textbf{y}^{(t+1)}_i- \bar{\textbf{s} }^{(t)}\right \|_1 \right \} .
\end{equation}

\noindent
For Eq. (\ref{lemma21eq_tem}), since at every $0\le k \le t-1$ the update $\textbf{s}_{i,noise}^{(k+\frac{1}{2})}-\textbf{s}_{i}^{(k)}$ is:
\begin{equation}\label{lemma21eq8}
\textbf{s}_{i,noise}^{(k+\frac{1}{2})}-\textbf{s}_{i}^{(k)} = \gamma_n\textbf{n}_i^{(k)} +\varepsilon ^{(k)}_{i},
\end{equation}

\noindent
and for iteration $t$,
\begin{equation}\label{lemma21eq9}
\textbf{s}_{i,noise}^{(t+\frac{1}{2})}-\textbf{s}_{i}^{(t)} = \textbf{s}_{i}^{(t+\frac{1}{2})}-\textbf{s}_{i}^{(t)} = \varepsilon ^{(t)}_{i}.
\end{equation}

\noindent
Following Theorem 1 in\cite{SGPL1}, we have
\begin{equation}\label{lemma21eq10}
\begin{aligned}
2 \left \|\textbf{y}^{(t+1)}_i- \bar{\textbf{s} }^{(t)}\right \|_1 \le& 2C'\lambda^{t}\left \|\textbf{s}_{i}^{(0)} \right \|_1 +2C'\left \|\varepsilon ^{(t)}_{i}\right \|_1\\
&+ 2 C'\sum_{k=0}^{t-1}\lambda^{t-k}\left \|\gamma_n\textbf{n}_i^{(k)} + \varepsilon ^{(k)}_{i} \right \|_1 .
\end{aligned}
\end{equation}

\noindent
Rearranging Eq. (\ref{lemma21eq10}), we obtain the conclusion.
\end{proof}

Based on the result in Lemma 2, we can derive the following iterative procedure that allows each node to efficiently compute its local sensitivity.

\noindent
\textbf{Remark 1}: {\it
For node $i\in [N]$ at time $t\ge 0$, the value of $S_i^{(t)}$ can be computed by the following iterative algorithm derived from Eq.~\eqref{SensPerIter}.
\begin{equation}
\label{iterSen}
S_i^{(t)} = 
\left\{
\begin{aligned}
& 2C'\left \| \textbf{s}_i^{(0)} \right \|_1 + 2C'\left \| \varepsilon ^{(0)}_{i} \right \|_1,\quad\quad\ \ t=0.\\
& \lambda S_i^{(t-1)} + 2C'\left (\left \| \varepsilon ^{(t)}_{i} \right \|_1 + \lambda \gamma_n\left \| \textbf{n}_{i}^{(t-1)} \right \|_1\right ),\\
&\quad\quad\quad\quad\quad\quad\quad\quad\quad\quad\quad\quad\quad\quad\quad t>0.
\end{aligned}
\right.
\end{equation}
\noindent
}

\noindent
As previously analyzed, the additional communication and computation overhead incurred by the DPPS protocol when using Eq.~\eqref{iterSen} for sensitivity estimation is acceptable. Regarding memory cost, each node $i\in [N]$ at time $t> 0$ only needs to store two scalars: the norm of the current round’s gradient and the norm of the previous round’s noise, to compute $S_i^{(t)}$ efficiently, introducing negligible additional memory overhead. Our subsequent theoretical analysis proves that, by using $S^{(t)}$ as the sensitivity, the mapping M, obtained by modifying the mapping m according to Lemma 1, is differentially private. This in turn establishes that the DPPS protocol satisfies differential privacy. Moreover, our later experiments validate the effectiveness of this sensitivity estimation by comparing the actual sensitivity observed during the execution of the DPPS protocol with the value $S^{(t)}$ computed via Eq.~\eqref{iterSen}.

\begin{algorithm}
\caption{Partial Communication Push-SUM SGD with Differential Privacy}
\textbf{Input}: Number of iterations $T$, local and shared learning rates $\gamma_l, \gamma_s$, noise rate $\gamma_n$, clipping threshold $\mathfrak{C}$, privacy budget hyperparameter $b$. Each node $i$ initializes its parameters $\textbf{s}_{i}^{(0)}, \textbf{l}_{i}^{(0)}$, corrected parameters $\textbf{y}_{i}^{(0)}=\textbf{s}_{i}^{(0)}$, and normalizing scalar $a^{(0)}_{i}=1$.\\
\textbf{Output}: $\bar{\textbf{s}}^{(T)}, \textbf{L}^{(T)}$.
\begin{algorithmic}[1] 
\FOR{$t=\{0, 1, ..., T-1\}$}
\FOR{each node $i\in [N]$ in parallel}

\STATE Randomly sample a batch of data $\xi ^{(t)}_{i}\sim D_{i}$.

\STATE Update the local parameters $\textbf{l}$ via:
\begin{equation}\label{localUpdate}
\textbf{l}_{i}^{(t+1)} = \textbf{l}_{i}^{(t)} - \gamma_l \textbf{g}_{i,l}^{(t)} ;
\end{equation}

\STATE Clip the partial derivative of the shared parameters $\textbf{s}$ for differential privacy via:
\begin{equation}\label{gradientclip}
\textbf{g}_{i,s}^{(t)}=\frac{\nabla _s F_i \left ( \textbf{y}_{i}^{(t)},\textbf{l}_{i}^{(t+1)};\xi ^{(t)}_{i} \right )}{\max \left ( 1,\left \| \nabla _s F_i \left ( \textbf{y}_{i}^{(t)},\textbf{l}_{i}^{(t+1)};\xi ^{(t)}_{i} \right ) \right \|_1 /\mathfrak{C}   \right ) };
\end{equation}

\STATE Update and communicate the shared parameters $\textbf{s}$ using the DPPS protocol with the perturbation $\varepsilon ^{(t)}_{i}$ as follows:
\begin{equation}\label{sharedUpdate}
\varepsilon ^{(t)}_{i} = -\gamma_s\textbf{g}_{i,s}^{(t)} ;
\end{equation}

\ENDFOR
\ENDFOR
\end{algorithmic}
\end{algorithm}

\subsection{PartPSP Algorithm}

Having established the DPPS protocol as a general-purpose private communication primitive, we now demonstrate its application in decentralized optimization. Specifically, we instantiate DPPS in the context of non-convex stochastic optimization to develop PartPSP, as outlined in Algorithm 2. In the PartPSP algorithm, each node maintains both local parameters and shared parameters. At every iteration $t$, each node $i\in[N]$ first randomly samples a data batch  $\xi ^{(t)}_{i}$ (Line 3), then updates its local parameters using the local gradient $ \textbf{g}_{i,l}^{(t)}$ (Line 4), and computes a clipped gradient $\textbf{g}_{i,s}^{(t)}$ of the shared parameters for the convention of differential privacy (Line 5). This clipped gradient is multiplied by the shared learning rate $\gamma_s$ to form the perturbation (Line 6), which is subsequently fed into the DPPS protocol to enable privacy consensus for the shared parameters. The output after iterations consists of the averaged shared parameters and the collection of local parameters.


An essential feature of PartPSP is its partial communication mechanism, which aims to achieve a better balance between privacy preservation and optimization efficiency. This strategy stems directly from an insight revealed by Lemma 2. Notably, the sensitivity upper bound given in Eq.~\eqref{SensPerIter}, derived from Lemma 2, may suffer from sensitivity explosion due to accumulated noise, which in turn leads to excessive noise injection and degrades PartPSP's optimization performance. More specifically, observing the right side of Eq.~\eqref{SensPerIter}, the first term $\lambda^t \left \|\textbf{s}_{i}^{(0)} \right \|_1$ converges to zero because $\lambda\in(0,1)$. The second term $\sum_{k=0}^{t}\lambda^{t-k}\left \|\varepsilon ^{(k)}_{i} \right \|_1$ asymptotically converges to a constant, because in PartPSP we have $\varepsilon ^{(k)}_{i} = -\gamma_s\textbf{g}_{i,s}^{(k)}$ and gradient clipping ensures $\left \|\textbf{g}_{i,s}^{(k)} \right \|_1\le \mathfrak{C}$, implying $\left \|\varepsilon ^{(k)}_{i} \right \|_1\le \gamma_s\mathfrak{C}$. However, the third term $\sum_{k=0}^{t-1}\lambda^{t-k}\left \|\textbf{n}_{i}^{(k)} \right \|_1 $ lacks a finite upper bound, potentially causing $S_i^{(t)}$ to diverge. We refer to this third term as the accumulated noise. Synchronizing the noised shared parameters $\textbf{s}_{i,noise}^{(t+\frac{1}{2})}$ across the network is the only way to eliminate the effect of accumulated noise, as it unifies all noised shared parameters and resets the sensitivity to zero. However, frequent synchronization undermines the low communication overhead advantage of decentralized learning. Therefore, controlling the impact of accumulated noise on sensitivity is essential to reduce the required synchronization frequency.

\begin{figure}[t]
    \centering
    
    \subfloat[Full communication]{%
        \includegraphics[width=0.475\linewidth, height=4cm, keepaspectratio]{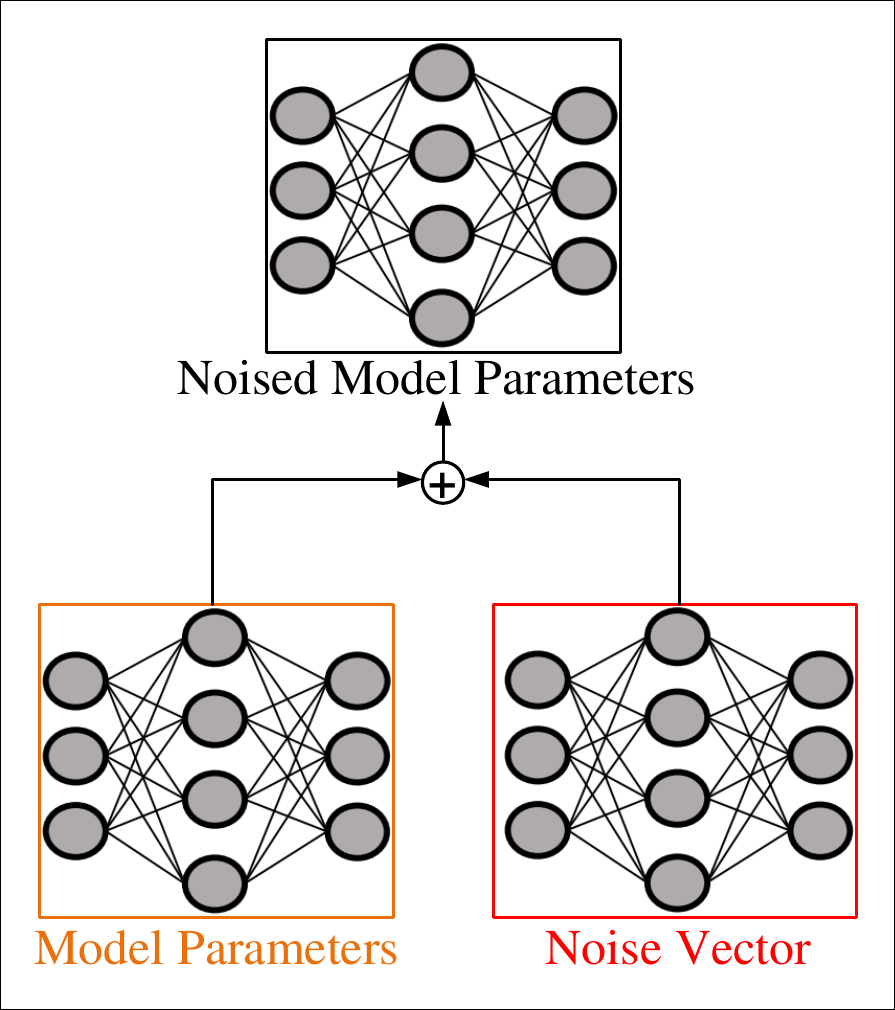}%
    }~
    \subfloat[Partial communication]{
        \includegraphics[width=0.475\linewidth, height=4cm, keepaspectratio]{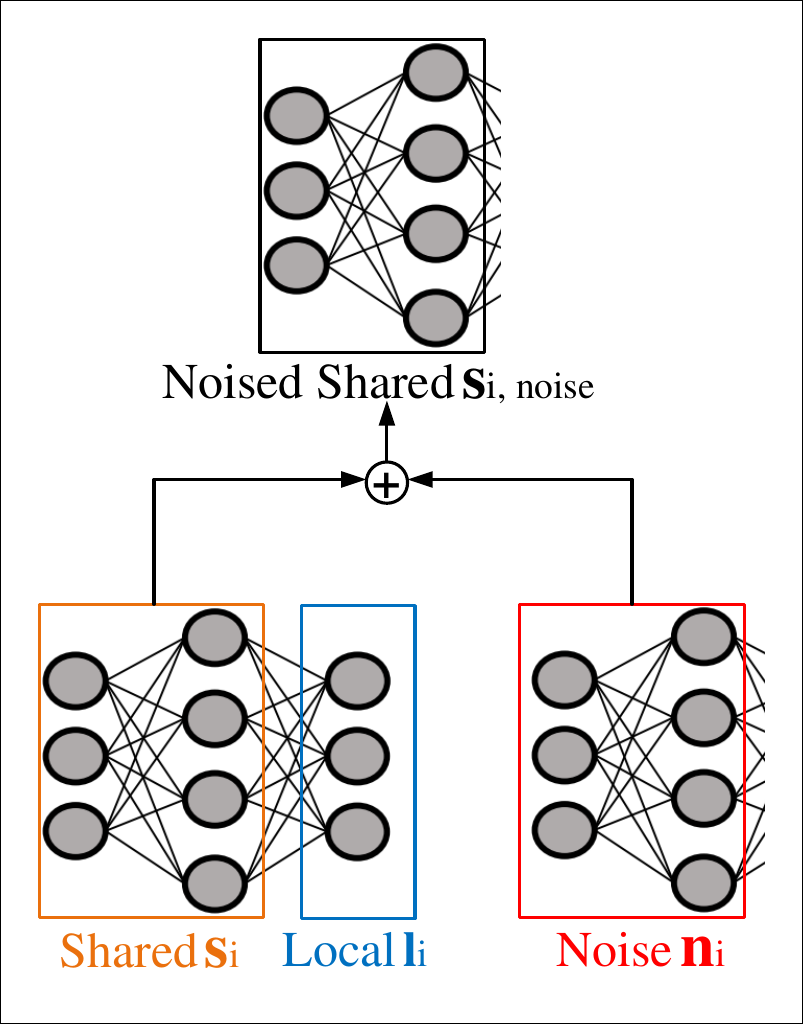}
    }
    
    \caption{Comparison of full and partial communication mechanisms.}
    \label{FullandPart}
\end{figure}

To mitigate the adverse effect of accumulated noise on sensitivity and thereby reduce the need for frequent synchronization, PartPSP adopts a partial communication mechanism. Although the underlying DPPS protocol adds independent Laplace noise to every entry of the shared parameters $\textbf{s}_i$, the magnitude of this noise accumulation depends directly on the dimension $d_s$. By design, PartPSP splits the model into the shared parameters $\textbf{s}_i$ and the local parameters $\textbf{l}_i$, communicating only $\textbf{s}_i$ via DPPS. This reduces $d_s$, which in turn lowers both the total amount of injected noise and the growth rate of the accumulated-noise term in Eq.~\eqref{SensPerIter}. Consequently, the sensitivity remains better controlled, allowing less frequent synchronization. Fig.~\ref{FullandPart} illustrates the contrast between full and partial communication. In the traditional full communication setting (Fig.~\ref{FullandPart}(a)), all model parameters are shared and thus perturbed by noise during DPPS execution. As the model size grows, so does the accumulated noise, leading to rapid growth of $S_i^{(t)}$ and requiring more frequent synchronization. In contrast, under partial communication (Fig.~\ref{FullandPart}(b)), only the shared subset $\textbf{s}_i$ is communicated and noised, while the local parameters $\textbf{l}_i$ remain private and unperturbed. This design effectively limits the dimensionality of the noise injection process and suppresses the divergence of sensitivity.

The partial communication mechanism is simple and clear, as it controls the sensitivity by adjusting the proportion of shared and local parameters in the model. However, whether PartPSP with partial communication converges for non-convex optimization problems becomes a critical problem that must be addressed. Later, we will provide a theoretical guarantee for the convergence of the PartPSP algorithm on non-convex problems. 

\section{Theoretical Guarantees}
\subsection{Security Guarantee}
We now provide the theoretical guarantee for the privacy protection offered by the DPPS protocol in the following theorem, which serves as the privacy-preserving backbone of PartPSP.

\noindent
{\bf Theorem 1}.{ \it
For each iteration $t>0$, let the noiseless version of the DPPS protocol be denoted by the mapping $m$. Then, the DPPS protocol that adds Laplace noise calibrated to the $L_1$ sensitivity $S^{(t)}$ satisfies $\frac{b}{\gamma_n}$-differential privacy.
}

\begin{proof} 
Consider the mapping $m$ in Definition 3 and its $L_1$ sensitivity in Definition 5. Viewing the noiseless version of the DPPS protocol as the mapping $m$, its $L_1$ sensitivity at each iteration $t>0$ during the execution is given by:

\begin{equation}\label{mL1}
\begin{aligned} 
\left \| m(q) - m(q') \right \|_1 &= \left \|\left [ \textbf{s}^{(t+\frac{1}{2})}  \right ] _{q} - \left [ \textbf{s}^{(t+\frac{1}{2})}  \right ] _{q'}\right \|_1\\
&\le \max_{i,j\in [N]}\left \|\textbf{s}^{(t+\frac{1}{2})}_i - \textbf{s}^{(t+\frac{1}{2})}_j \right \|_1 \\
&\le \max_{i\in [N]}\left \{ S_i^{(t)} \right \} \\
&\le S^{(t)},
\end{aligned}
\end{equation}

\noindent
where queries $q, q' \in Q$ differ in only one entry. In the DPPS protocol, the output parameter to the query $q$ is the noised shared parameter $\textbf{s}_{noise}^{(t+\frac{1}{2})}$. In other words, the DPPS protocol constructs a mapping $M$ based on $m$, which is formalized as follows:

\begin{equation}\label{mapM}
M(q) = m(q) + \gamma_n L_{noise}^{(t)},
\end{equation}
where $L_{noise}^{(t)}$ consists of elements i.i.d. according to the Laplace distribution $Lap(0, \frac{S^{(t)}}{b})$. 

Based on Lemma 1 and the properties of the Laplace distribution, the DPPS protocol satisfies $\frac{b}{\gamma_n}$-differential privacy at each iteration $t>0$.
\end{proof}


\subsection{Convergence Guarantee}
We now discuss the convergence guarantee of PartPSP for non-convex objective functions. As described in the Preliminaries, the objective of PartPSP is to find solutions to Problem 1, which is a unified shared parameters $\textbf{s}$ and a set $\textbf{L}$ of local parameters. Conventional algorithms based on the Push-SUM protocol require an upper bound on the gradient of the non-convex objective functions with respect to the model's parameters. In this paper, we divide the model's parameters into two parts: the shared parameters $\textbf{s}$ and the local parameters $\textbf{l}$, and optimize them using their respective partial derivatives. Therefore, we need to provide new indicators to estimate the optimization performance of PartPSP for $\textbf{s}$ and $\textbf{L}$. We follow the previous work \cite{DFedPGP} in personalized federated learning to provide the following definitions of partial derivatives.

\noindent
{\bf Definition 8}.{ \it
We define the following two partial derivatives as a measure of how closely the decentralized network approaches the solution to Problem 1.
\begin{equation*}
\Delta^{(t)}_{\bar{\textbf{s} } }=\mathbb{E}\left [  \left\|\nabla _s F\left ( \bar{\textbf{s} }^{(t)},\textbf{L}^{(t+1)} \right )\right \|_2^2\right ],
\end{equation*}
\begin{equation*}
\Delta^{(t)}_{\textbf{l} }=\frac{1}{N}\sum_{i=1}^{N}  \mathbb{E}\left [   \left \| \nabla _l F_i\left ( \textbf{y} ^{(t)}_{i},\textbf{l}^{(t)}_{i}\right ) \right \| _2^2\right ].
\end{equation*}
}

\noindent
$\Delta^{(t)}_{\bar{\textbf{s} } }$ describes the average partial derivative of the network with respect to the unified shared parameters $\bar{\textbf{s}}^{(t)}$ without considering the set $\textbf{L}^{(t)}$ of local parameters. $\Delta^{(t)}_{\textbf{l} }$ describes the average local partial derivative of the network with respect to each node's local parameters, which compose the set $\textbf{L}^{(t)}$.

On the other hand, since PartPSP provides privacy protection for each node $i \in [N]$ through the differentially private DPPS protocol, the shared parameter $\textbf{s}_i$ is perturbed with noise before communication. At iteration $t$, the magnitude of the injected noise is determined by the sensitivity $S^{(t)}$. However, the explicit expression for $S^{(t)}$ given in Lemma 2 is too complex to facilitate convergence analysis. Therefore, we introduce the following sensitivity assumption to simplify convergence analysis:

\noindent
{\bf Assumption 4}.{ \it
There exists a constant $S > 0$ such that $S^{(t)} \leq S$ for all $t \geq 0$ in PartPSP, where $S$ is the assumed sensitivity upper bound for PartPSP.
}

We present the following theorem that establishes the convergence guarantee of PartPSP for optimizing non-convex objective functions.

\begin{figure*}[t]
    \centering
    \subfloat[MLP, 2-Out, 1 layer shared]{%
        \includegraphics[width=0.245\linewidth]{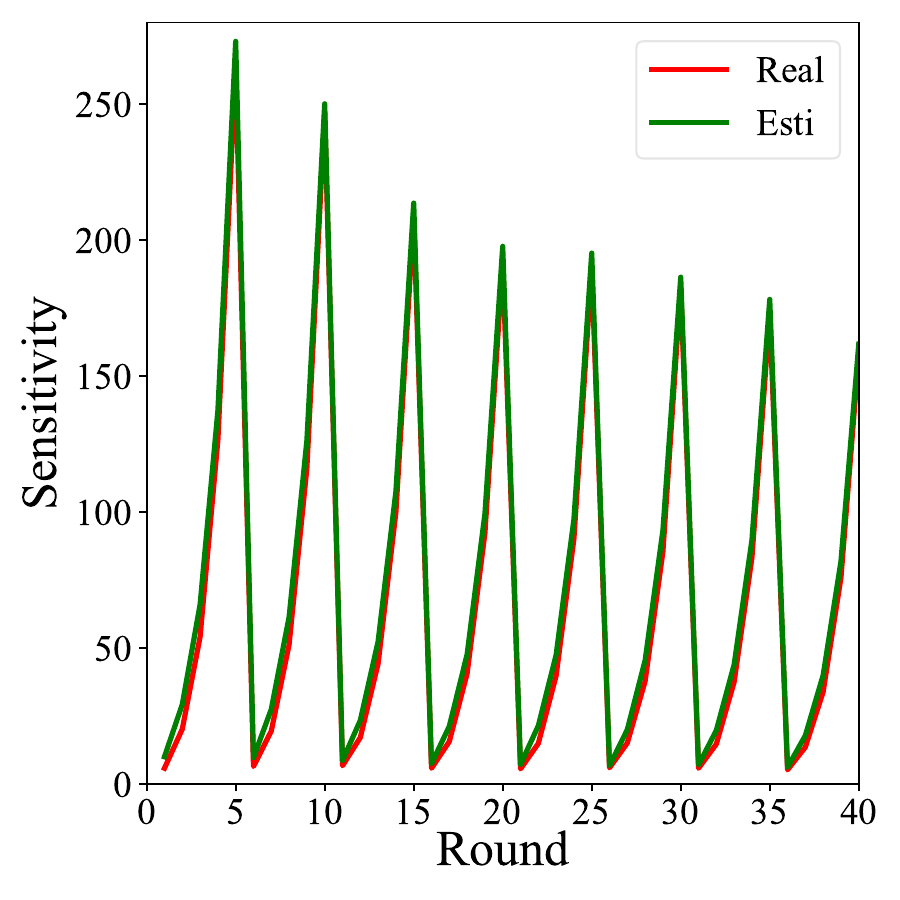}%
    }~
    \subfloat[MLP, 2-Out, 2 layers shared]{%
        \includegraphics[width=0.245\linewidth]{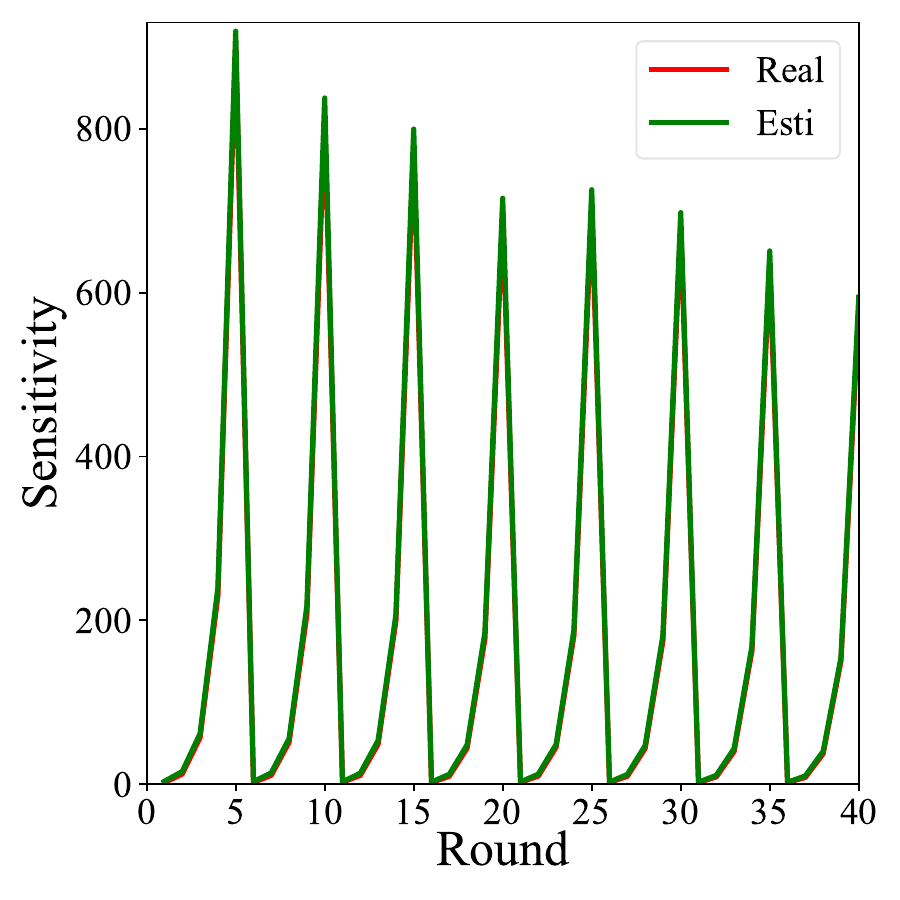}%
    }
    \subfloat[MLP, Exp, 1 layer shared]{%
        \includegraphics[width=0.245\linewidth]{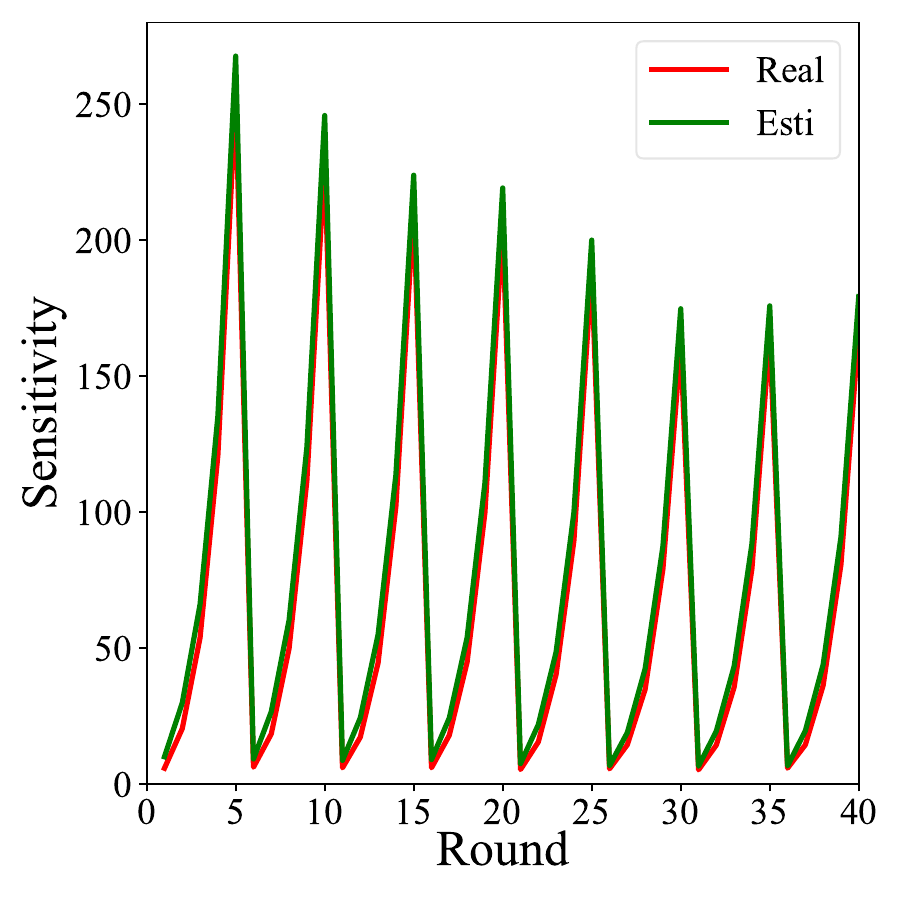}%
    }~
    \subfloat[MLP, Exp, 2 layers shared]{%
        \includegraphics[width=0.245\linewidth]{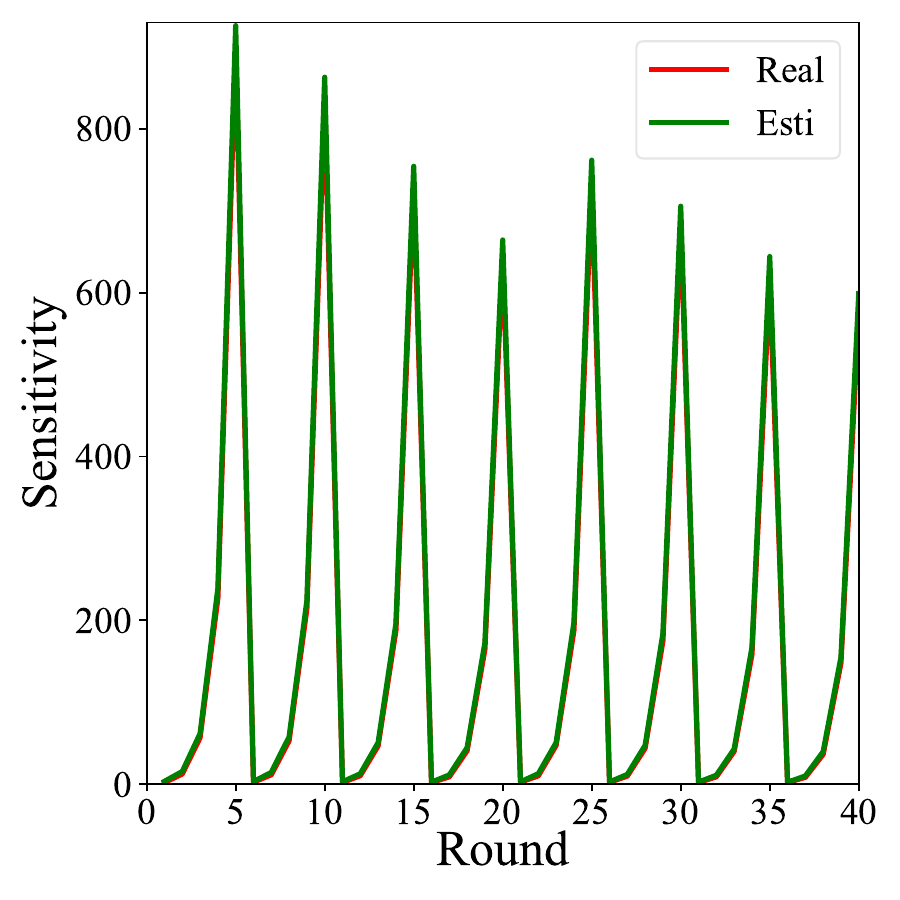}%
    }

    \subfloat[ResNet, 2-Out, 1 layer shared]{%
        \includegraphics[width=0.245\linewidth]{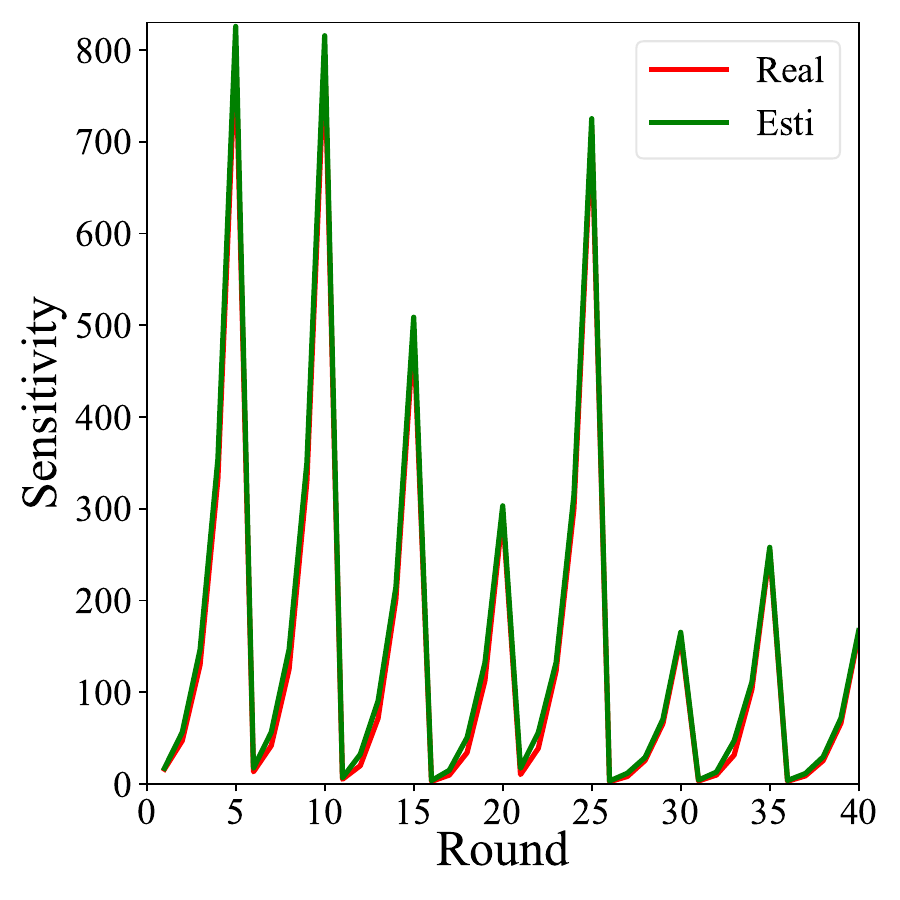}%
    }~
    \subfloat[ResNet, 2-Out, 2 layers shared]{%
        \includegraphics[width=0.245\linewidth]{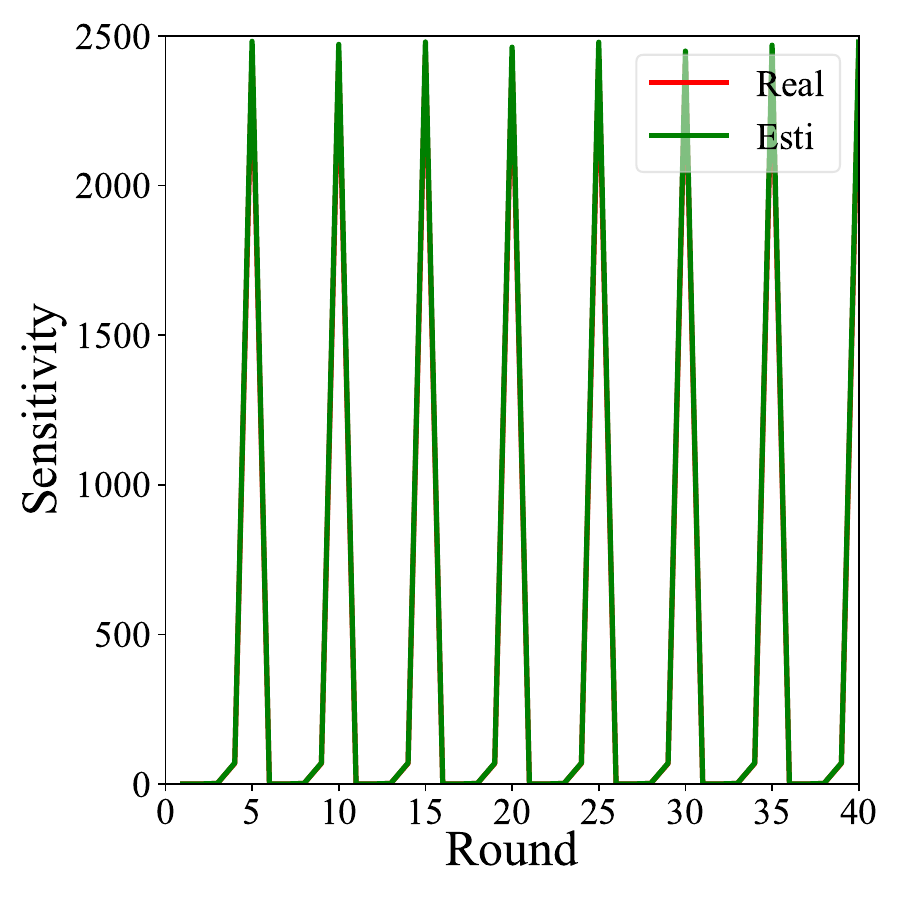}%
    }
    \subfloat[ResNet, Exp, 1 layer shared]{%
        \includegraphics[width=0.245\linewidth]{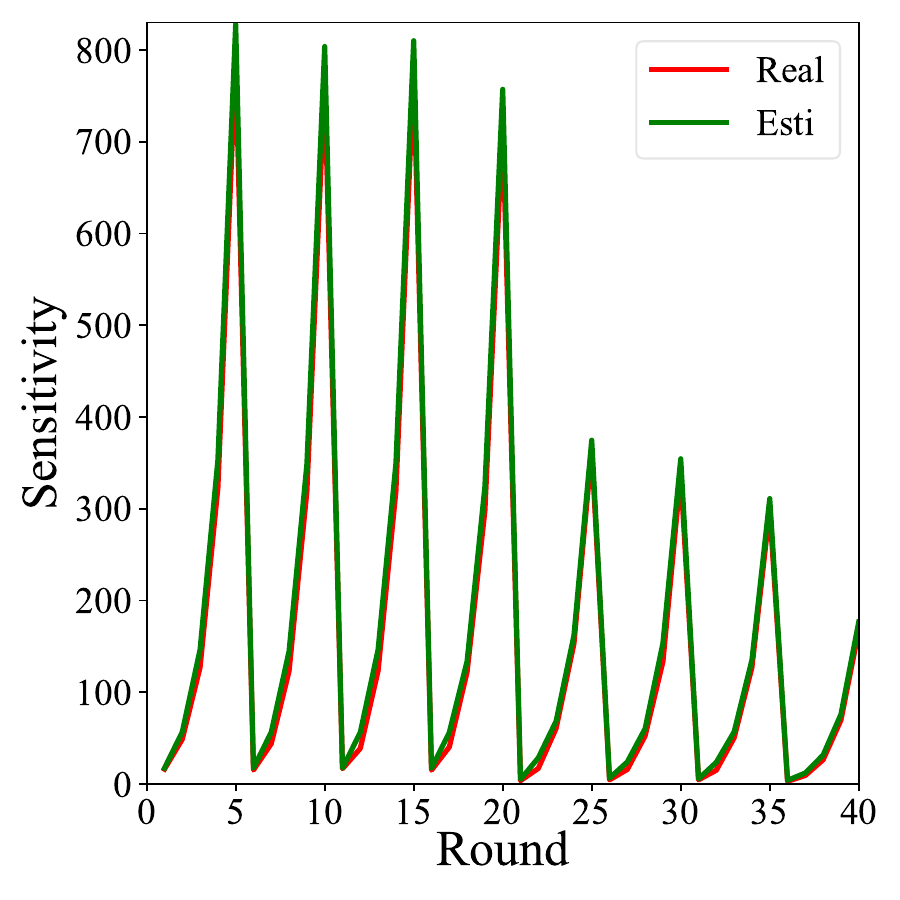}%
    }~
    \subfloat[ResNet, Exp, 2 layers shared]{%
        \includegraphics[width=0.245\linewidth]{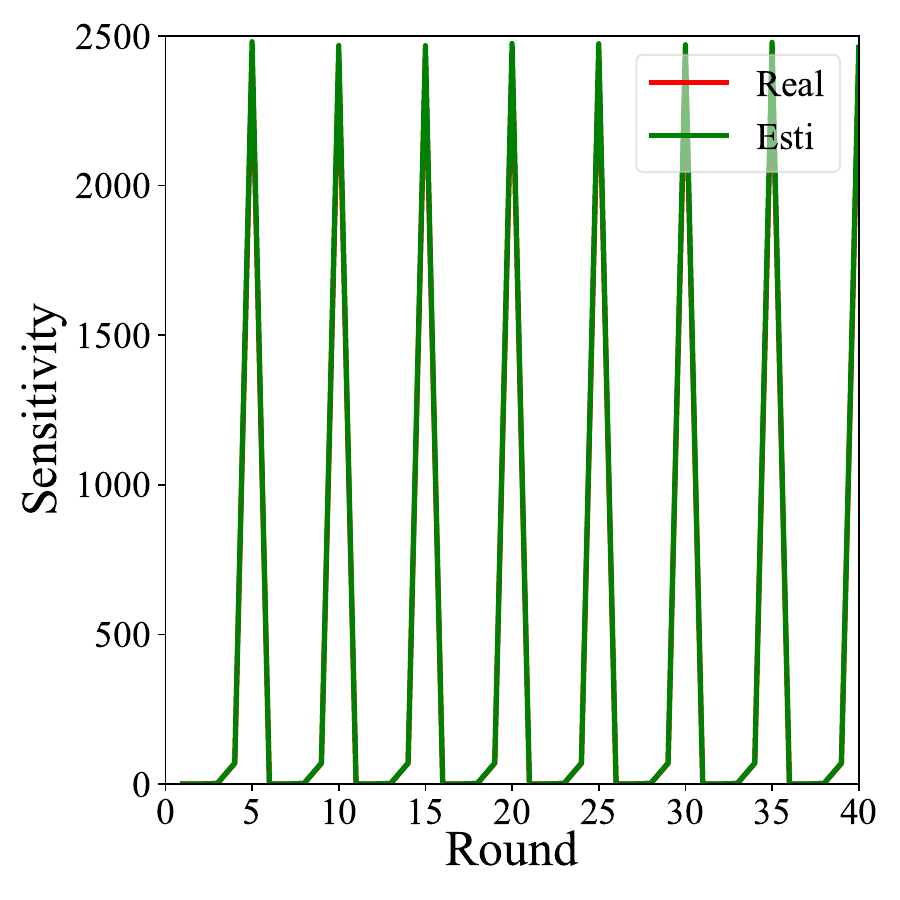}%
    }
    \caption{Comparison of Real Sensitivity and Estimated Sensitivity.}
    \label{Real_Esti_Sen}
\end{figure*}

\noindent
{\bf Theorem 2}.{ \it
Suppose that Assumptions 1, 2, 3, and 4 hold and that the weight matrix satisfies Definition 1. Under step-sizes $\gamma_s <\min\left \{  \frac{1-q}{2\sqrt{15} CL_{sc}},\frac{1}{16L_s+8A}  \right \} $, $\gamma_l < \frac{1}{1+L_l} $, and $\gamma_n < \frac{\gamma_s}{2}$, and satisfying $\gamma_l=O\left ( \frac{1}{\sqrt{T} }  \right )$, $\gamma_s=O\left ( \frac{1}{\sqrt{T} }  \right )$, and $\gamma_n=O\left ( \frac{1}{T}  \right )$. For PartPSP, when $T$ is sufficiently large, we have the following conclusion.
\begin{equation*}
\begin{aligned}
&&& O\left ( \frac{1}{T}\sum_{t=0}^{T-1}\Delta^{(t)}_{\textbf{l} }\right )+O\left ( \frac{1}{T} \sum_{t=0}^{T-1}\Delta^{(t)}_{\bar{\textbf{s} } }\right )\\ 
&&& \le O\left ( \frac{1}{\sqrt{T}} \right ) +O\left ( \frac{d_s S^2}{b^2\sqrt{T}} \right ) \\
&&&\quad +O\left ( \frac{\sigma_l^2}{\sqrt{T}} \right ) +O\left ( \frac{\sigma_s^2}{\sqrt{T}} \right ) +O\left ( \frac{\sigma_g^2}{\sqrt{T}} \right ) .
\end{aligned}
\end{equation*}
\noindent
The proof of Theorem 2 is available in the Appendix.
}

Theorem 2 provides convergence guarantees for both local parameters $\textbf{l}$ and shared parameters $\textbf{s}$ under non-convex conditions, using an upper bound on the sum of their partial derivatives. The second term $O\left ( \frac{d_s S^2}{b^2\sqrt{T}}\right )$ in the upper bound of Theorem 2 captures the theoretical degradation to PartPSP's optimization performance caused by the differential privacy mechanism of the DPPS protocol. On the one hand, reducing the dimension $d_s$ of shared parameters can directly decrease the value of this term. On the other hand, Lemma 2 shows that the sensitivity upper bound $S$ (Assumption 4) is positively correlated to $d_s$. Therefore, PartPSP can reduce its dimension $d_s$ of shared parameters to decrease the gradient term $O\left ( \frac{d_s S^2}{b^2\sqrt{T}}\right )$, thereby to increasing its optimization performance. The trade-off between privacy protection and optimization performance is an inherent contradiction in every privacy-preserving algorithm that achieves privacy through differential privacy. According to Theorem 2, PartPSP can theoretically improve its optimization performance after introducing differential privacy by reducing the dimension $d_s$ of shared parameters via partial communication. Therefore, partial communication endows PartPSP with a unique capability to reduce the performance degradation caused by the differential privacy.
\section{Experiments}
\subsection{Experimental Settings}
Our experiments are based on three commonly used datasets: MNIST \cite{MNIST}, FMNIST \cite{FMNIST}, and CIFAR-10 \cite{CIFAR}. For all datasets, we use the DistributedSampler from PyTorch as the data sampler, and the batch size at each node is set to 100.

For the MNIST dataset, we employ a Multi-Layer Perceptron (MLP) model with three linear layers. The architecture of the MLP is as follows: the first layer is a $784 \times 10$ linear layer, followed by a $10 \times 784$ linear layer, and then another $784 \times 10$ linear layer. Each layer has 7840 parameters. The activation function between each layer is the hyperbolic tangent (Tanh), which is a smooth function. For the FMNIST and CIFAR-10 datasets, we utilize the ResNet-18 \cite{Resnet} and ViT \cite{ViT} models, respectively. These tested models include Convolutional Neural Network and Transformer architectures, which represent the mainstream deep learning models.

If not specifically pointed out, we set the number of nodes to $N=10$, use Python 3.12, PyTorch 2.4, and CUDA 12.4, and put all random seeds to 2024. Our experiments are based on the graphs called $d$-Out and EXP \cite{SGP}.

\noindent
\textbf{Remark 2}: {\it
The $d$-Out and EXP graphs are two directed decentralized network graphs. In the $d$-Out graph, the outdegree of every node is $d$. During each communication round, any node $i \in [N]$ sends its shared parameters to the nodes indexed from $(i+0) \mod N$ to $(i+d-1) \mod N$. In the EXP graph, each node $i \in [N]$ periodically sends its shared parameters to the node indexed by $(i + 2^{(t \mod (\lfloor \log_2(N-1) \rfloor + 1))}) \mod N$.
}

\noindent
In our experiments, both the $d$-Out and EXP graphs are directed. Specifically, the EXP graph is time-varying and is a widely used decentralized network setting for the Perturbed Push-SUM protocol. To ensure that the weighting matrix satisfies Definition 1, we assign each node $i\in [N]$ a weight of $\frac{1}{d_{i,out}}$ to each of its out-neighbors, where $\frac{1}{d_{i,out}}$ denotes its out-degree. In the $d$-Out graph, every node has exactly $d$ out-neighbors, so the assigned weight is $\frac{1}{d}$. In the EXP graph, each node maintains exactly two out-neighbors at every round, resulting in a uniform weight of $\frac{1}{2}$. The resulting weighting matrices under this construction are doubly stochastic, thus satisfying the requirements in Definition 1.

\subsection{Comparison of Real vs. Estimated Sensitivity}
The privacy guarantee of the DPPS protocol at the protocol level hinges on a critical requirement: the \textbf{estimated sensitivity} $S^{(t)}$ used for adding noise must be greater than or equal to the \textbf{real sensitivity} $\max_{i,j\in [N]}\left \|\textbf{s}^{(t+\frac{1}{2})}_i - \textbf{s}^{(t+\frac{1}{2})}_j \right \|_1$. To validate the effectiveness of DPPS's privacy mechanism, we leverage the downstream application PartPSP and compare the estimated sensitivity with the real sensitivity observed during optimization, summarizing the results in Fig.~\ref{Real_Esti_Sen}. Fig.~\ref{Real_Esti_Sen} illustrates the comparison of estimated and real sensitivity for PartPSP when training MLP and ResNet-18 models on the MNIST and FMNIST datasets, with either 1 or 2 layers of model parameters shared, under the 2-Out graph and EXP graph. For all experiments in Fig.~\ref{Real_Esti_Sen}, the privacy budget hyperparameter $b=5$, the constant $C'$ in Eq.\eqref{SensPerIter} is set to 0.78, and the synchronization interval is set to 5. For the MLP, the constant $\lambda$ for sharing one layer is set to 0.55, and for sharing two layers it is set to 0.62. For the ResNet-18, $\lambda$ for sharing one layer is set to 0.53, and for sharing two layers it is set to 0.7, with the gradient clipping threshold $\mathfrak{C}$ set to 100.

The Real and Esti in Fig. \ref{Real_Esti_Sen} represent the real and estimated sensitivity at each round, respectively. All Esti curves are strictly above the Real curves in Fig. \ref{Real_Esti_Sen}, which indicates that the DPPS protocol, used within PartPSP, provides valid privacy protection in all experiments shown. As we can see, the real sensitivity during the execution of PartPSP is almost identical to the estimated sensitivity calculated from Eq.\eqref{SensPerIter}. Therefore, the sensitivity estimation mechanism introduced for PartPSP is both theoretically sound and empirically accurate. 

By comparing the results that use the same models and network graphs but different shared layers, taking (a) and (b) in Fig. \ref{Real_Esti_Sen} as an example, we can see that the partial communication mechanism effectively reduces the peak sensitivity of PartPSP when optimizing the MLP from over 800 to below 300. Note that the number of parameters in each layer of the MLP is 7840. Thus, the reduction in sensitivity achieved by the partial communication mechanism does not scale linearly with the number of shared parameters. 

By comparing the results that use the same models and shared layers but different network graphs, such as (a) and (c) in Fig. \ref{Real_Esti_Sen}, we observe that the 2-Out and EXP graphs exhibit similar sensitivity trends in most cases. Note that in the corresponding experiments of PartPSP on the 2-Out and EXP graphs, they use the same constants $C'$ and $\lambda$, and each node in both graphs has an in-degree and out-degree of 2 during each communication. Therefore, we conjecture that the influence of network structure on sensitivity is primarily determined by node degree. Subsequent experiments support this hypothesis.

\subsection{Factors Affecting Sensitivity}

Since the remaining experiments have thousands of communication rounds, we use the \textbf{real average sensitivity} (RAS) as an alternative to indicate the real sensitivity during an experiment. The real sensitivities of the remaining experiments are strictly lower than their estimated sensitivities, which means they are all under DPPS's privacy protection, and the detailed values are available in our open-source code.

\begin{figure}[t]
    \centering
    \subfloat[RAS about shared layers]{%
        \includegraphics[width=0.475\linewidth]{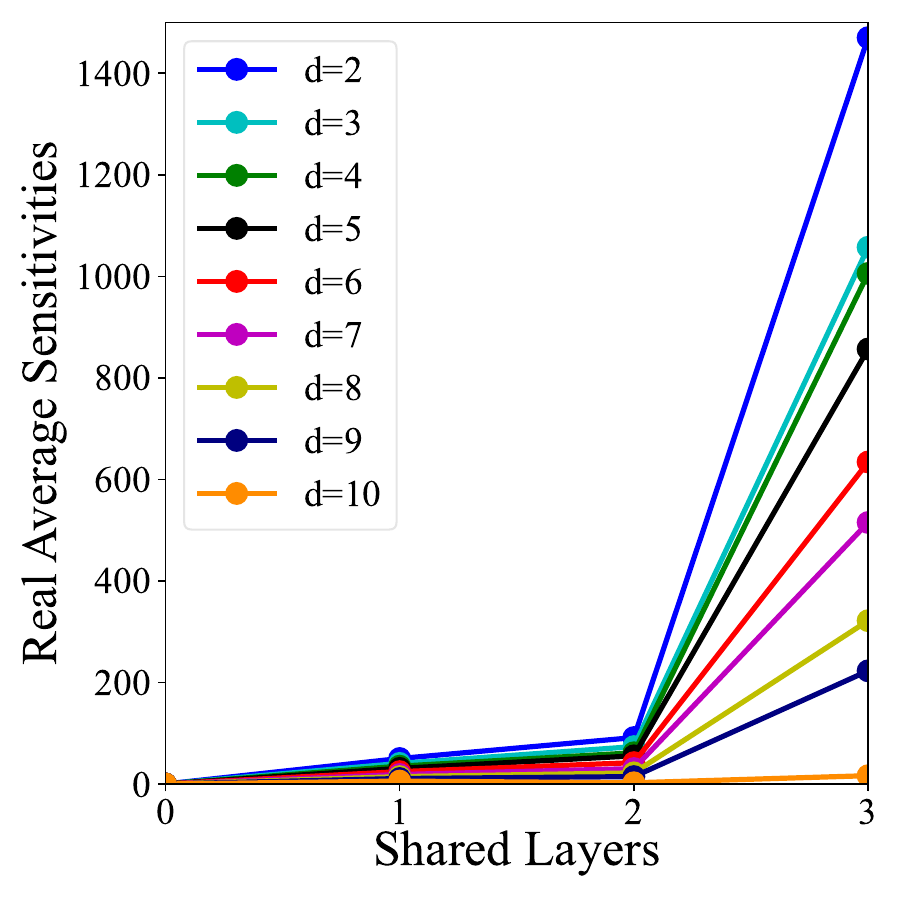}%
    }~
        \subfloat[RAS about $d$-Out graphs]{%
        \includegraphics[width=0.475\linewidth]{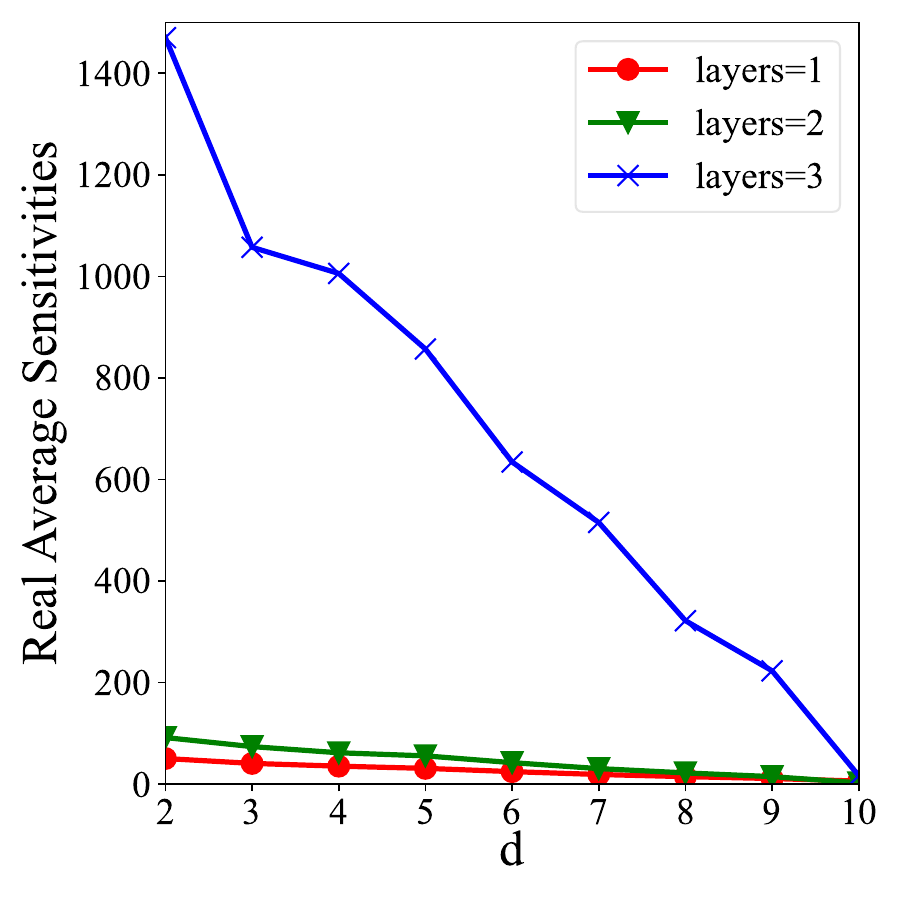}%
    }
    \caption{RAS about partial communication and network connectivity.}
    \label{RASOutLay}
\end{figure}

Having verified the validity of the estimated sensitivity provided by Eq.~\eqref{SensPerIter} for the DPPS protocol, we further observe that, according to the same expression, the sensitivity during DPPS execution is also influenced by network connectivity. Specifically, this influence is mediated through the constants $C'$ and $\lambda$ in Eq.~\eqref{SensPerIter}, which control the scaling and decay of sensitivity over iterations. Theoretically, for two networks of the same scale, a larger node degree leads to smaller values of both $C'$ and $\lambda$, and consequently results in a lower estimated sensitivity. On the other hand, as a downstream instantiation of DPPS, PartPSP controls sensitivity through its partial communication mechanism. By reducing the dimensionality $d_s$ of the shared parameters, this mechanism yields a theoretical influence on the sensitivity incurred during DPPS execution. We aim to investigate how the above factors influence the sensitivity of DPPS during its execution. To this end, we conduct experiments based on PartPSP and report the observed RAS values under different partial communication and network connectivity settings in Fig.~\ref{RASOutLay}.

All the experiments in Fig. \ref{RASOutLay} use PartPSP to optimize MLP, the privacy budget hyperparameter $b=5$, the constants $C'=0.95$ and $\lambda=0.55$, and the synchronization interval is set to 4. Observing Fig. \ref{RASOutLay}(a), we can see that in all network connectivity settings, partial communication performs the expected effect: as the dimension $d_s$ of shared parameters decreases, its RAS also decreases. Note that although the number of parameters in each layer of the MLP is the same, the RAS does not decrease linearly but rather at a faster rate. This is because reducing $d_s$ through partial communication not only affects the dimension of $\textbf{n}$ but also influences the noise intensity of each element in $\textbf{n}$. The combined effect of these two factors causes the RAS of PartPSP to decrease much faster than the linear reduction of $d_s$. Fig. \ref{RASOutLay}(b) shows the same experimental results as (a) but from a different perspective. We observe that under all partial communication settings, the RAS of PartPSP decreases as the network connectivity of the $d$-Out graph increases. Therefore, the results in Fig. \ref{RASOutLay} observed during the practical execution of PartPSP verify our theoretical analysis regarding the influences of partial communication and network connectivity on the sensitivity of DPPS.

\begin{figure}[htbp]
    \centering
    \subfloat[RAS of Training MLP]{%
        \includegraphics[width=0.475\linewidth]{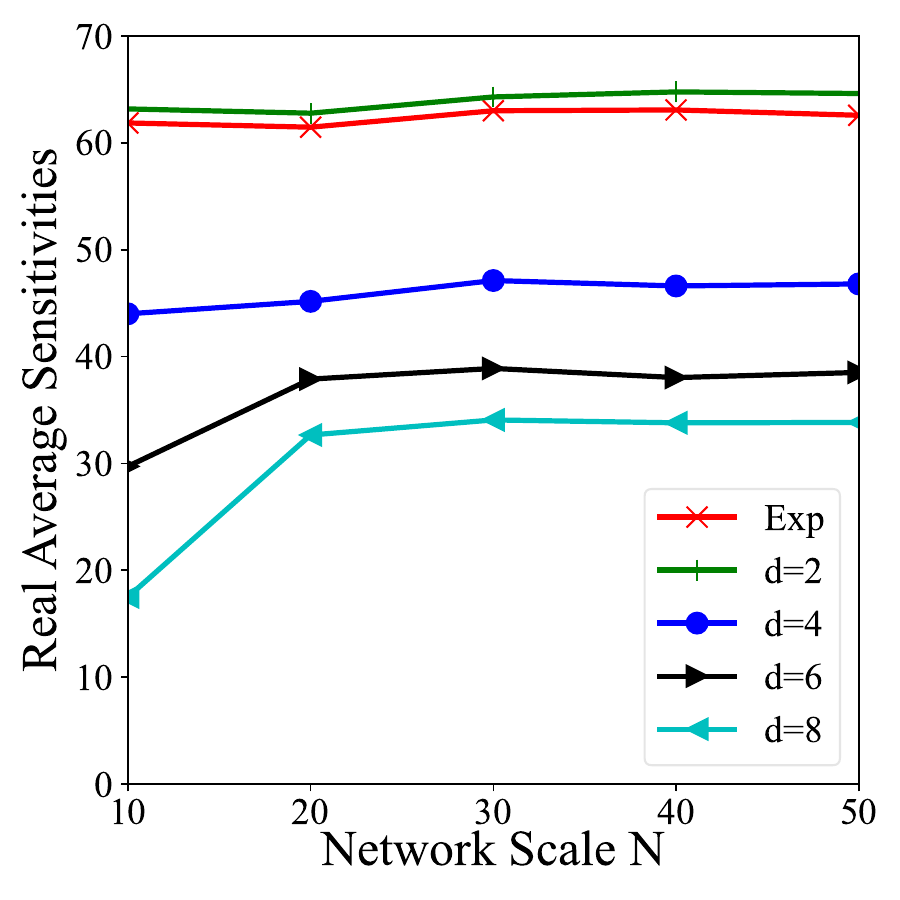}%
    }~
        \subfloat[RAS of Training ResNet-18]{%
        \includegraphics[width=0.475\linewidth]{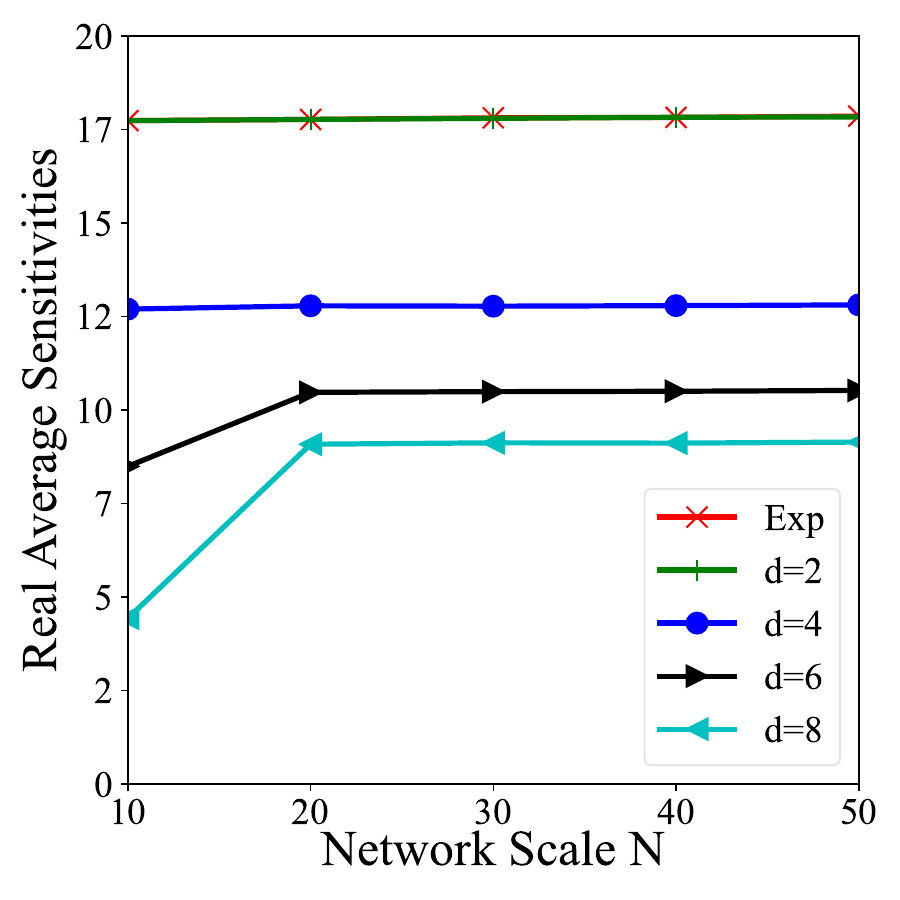}%
    }
    \caption{RAS about network scale.}
    \label{RAS_N}
\end{figure}

\begin{table*}[htbp]
\centering
\makeatletter
\makeatother
\caption{The Final Test Accuracy (\%)}
\begin{tabular}{c|c|cccc|cccc|cccc}
\toprule
\multicolumn{2}{c|}{Model/Dataset} & \multicolumn{4}{c|}{MLP/MNIST} & \multicolumn{4}{c|}{ResNet-18/FMNIST} & \multicolumn{4}{c}{ViT/CIFAR-10} \\
\cmidrule(lr){1-2} \cmidrule(lr){3-6} \cmidrule(lr){7-10} \cmidrule(lr){11-14}
\multicolumn{2}{c|}{Topology} & EXP & 4-Out & 6-Out & 8-Out & EXP & 4-Out & 6-Out & 8-Out & EXP & 4-Out & 6-Out & 8-Out \\
\midrule
\multirow{4}{*}{\centering PartPSP-1} & b=1.0 & \textbf{41.57} & \textbf{40.63} & 38.78 & 43.19 & \textbf{82.70} & \textbf{82.48} & \textbf{82.97} & \textbf{82.63} & \textbf{33.60} & \textbf{37.00} & \textbf{45.79} & \textbf{66.97} \\
                              & b=2.0 & \textbf{44.67} & \textbf{46.25} & \textbf{60.03} & \textbf{81.00} & \textbf{82.85} & \textbf{82.54} & \textbf{82.10} & \textbf{81.90} & \textbf{51.63} & \textbf{64.33} & \textbf{74.97} & \textbf{76.76} \\
                              & b=3.0 & \textbf{48.08} & \textbf{63.51} & \textbf{80.12} & \textbf{88.87} & \textbf{82.73} & \textbf{81.86} & 82.06 & \textbf{82.59} & \textbf{71.78} & \textbf{76.24} & \textbf{76.72} & \textbf{76.98} \\
                              & NoDP & \textbf{89.66} & \textbf{89.71} & \textbf{89.74} & \textbf{89.75} & 88.01 & 88.17 & 88.01 & 88.14 & 76.98 & 77.04 & 77.03 & 77.00 \\
\midrule
\multirow{4}{*}{\centering PartPSP-2} & b=1.0 & 21.66 & 25.42 & 28.09 & 22.03 & 69.14 & 81.13 & 80.95 & 81.65 & 18.55 & 18.41 & 17.33 & 22.10 \\
                              & b=2.0 & 23.33 & 24.78 & 25.70 & 28.68 & 80.17 & 81.11 & 81.29 & 80.93 & 21.49 & 24.74 & 28.10 & 49.24 \\
                              & b=3.0 & 18.84 & 24.07 & 34.02 & 67.26 & 81.60 & 80.75 & \textbf{82.22} & 81.37 & 29.47 & 40.79 & 54.65 & 71.34 \\
                              & NoDP & 88.85 & 88.91 & 88.95 & 88.99 & 87.33 & 87.49 & 87.37 & 87.42 & 78.45 & \textbf{78.54} & \textbf{78.50} & \textbf{78.48} \\
\midrule
\multirow{4}{*}{\centering SGPDP} & b=1.0 & 29.23 & 26.68 & 33.23 & 22.38 & 66.38 & 72.36 & 72.58 & 72.85 & 10.46 & 10.39 & 11.91 & 8.70 \\
                          & b=2.0 & 24.89 & 32.39 & 25.91 & 28.31 & 71.77 & 73.26 & 72.65 & 72.63 & 10.86 & 7.61 & 14.39 & 41.08 \\
                          & b=3.0 & 29.78 & 26.94 & 29.34 & 45.63 & 71.65 & 72.12 & 71.01 & 72.83 & 19.43 & 29.55 & 51.64 & 68.96 \\
                          & NoDP & 81.55 & 81.64 & 81.62 & 81.62 & \textbf{90.18} & \textbf{90.31} & \textbf{89.98} & \textbf{90.16} & \textbf{78.65} & 78.47 & 78.48 & 78.46 \\
\midrule
\multirow{4}{*}{\centering PEDFL} & b=1.0 & 38.47 & 34.81 & \textbf{44.72} & \textbf{44.37} & 31.32 & 38.71 & 55.81 & 49.09 & 9.27 & 10.28 & 10.80 & 9.83 \\
                          & b=2.0 & 42.94 & 43.55 & 44.41 & 44.16 & 53.66 & 50.11 & 51.97 & 71.16 & 11.71 & 12.27 & 15.40 & 17.34 \\
                          & b=3.0 & 44.17 & 44.40 & 44.24 & 44.22 & 51.30 & 48.80 & 49.43 & 73.44 & 24.93 & 32.18 & 36.24 & 45.79 \\
                          & NoDP & 84.36 & 84.36 & 84.31 & 84.25 & 72.53 & 72.22 & 71.65 & 72.61 & 75.46 & 74.17 & 75.48 & 77.37 \\
\bottomrule
\end{tabular}
\label{ACCLOSS}
\end{table*}

For large-scale decentralized networks, determining the sensitivity hyperparameters $C'$ and $\lambda$ for Eq.~\eqref{SensPerIter} can be very costly. To address this issue, we present in Fig. \ref{RAS_N} the impact of network scale $N$ and the graph $d$-Out on RAS during the execution of PartPSP. We observe that, except for some cases where $d$ is very close to $N$, such as $d = 6$ or $8$ when $N=10$, for other cases, the same $d$ exhibits similar RAS behavior across different network scales. This phenomenon is observed not only in experiments with MLPs but also in experiments with different ResNets. Therefore, we conclude that for a fixed node degree $d$, if $d$ is much smaller than the network scale, the sensitivity of DPPS in smaller networks is consistent with that in larger networks. Consequently, for large-scale decentralized networks, the sensitivity hyperparameters $C'$ and $\lambda$ for Eq.~\eqref{SensPerIter} can be set using configurations derived from smaller-scale networks, thereby significantly reducing the cost of hyperparameter selection.

\subsection{Optimization Performance}
To verify both the optimization capability of PartPSP and the effectiveness of its partial communication mechanism, we experimentally compared the performance of PartPSP using different partial communication strategies with several other algorithms on the MNIST, FMNIST, and CIFAR datasets. The algorithms tested in the experiments are listed as follows:

\noindent
{
\begin{itemize}
\item \texttt{PartPSP-1}: PartPSP shares the first MLP layer, the first ResNet convolutional block, and the ViT patch embedding layers.
\item \texttt{PartPSP-2}: PartPSP shares the first two MLP layers, the first two ResNet convolutional blocks, and the ViT patch embedding layers and encoders.
\item \texttt{SGPDP}: SGP \cite{SGP} with differential privacy, which fully communicates all model parameters.
\item \texttt{PEDFL} \cite{PEDFL}: A decentralized federated learning algorithm with differential privacy by Laplace mechanism.
\end{itemize}
}

In this part of the experiment, the sensitivity of all algorithms during execution is set to real sensitivity at the corresponding communication. This setting eliminates the possibility of distorting the experimental results through hyperparameters $C'$ and $\lambda$, and ensures the effectiveness of privacy protection across all experiments. We set all experiments' training epochs to 12 and tested the model's performance after training. For testing, we first collect all the shared parameters in the network to get the average shared parameters $\bar{\textbf{s}}$, then send it to every node. After receiving $\bar{\textbf{s}}$, each node tests the average shared parameters $\bar{\textbf{s}}$ and its local parameters $\textbf{l}_i$ (if the algorithm has them) on the entire test dataset to obtain its test accuracy. Finally, we collect all the test accuracies from each node and compute the average test accuracy as the final test accuracy of our result. All experiments on the algorithms used the same set of hyperparameters, which can be found in our open-source code. Table \ref{ACCLOSS} shows the performance of four algorithms under varying networks and privacy budgets. We bold the best optimization results among the four algorithms under the same privacy budget and network.

As shown in Table \ref{ACCLOSS}, PartPSP-1 wins 33 out of the 36 privacy-preserving experiments (b = 1, 2, 3), underperforming only against PEDFL in two MLP experiments and against its variant, PartPSP-2, in one ResNet-18 experiment. Comparing PartPSP with PEDFL, under privacy-preserving setups (b = 1, 2, 3), PartPSP-1 outperforms PEDFL in 34 of 36 experiments, while PartPSP-2 outperforms PEDFL in 25 of 36 experiments. The experiments won by PEDFL are all optimizing MLP, while PartPSP-1 and PartPSP-2 win all ResNet and ViT experiments. According to the results in Table \ref{ACCLOSS}, we believe the DPPS-based PartPSP has better scalability in optimizing large deep learning models based on CNN (ResNet) and Transformer (ViT) architectures. In the 12 non-private experiments (NoDP), PartPSP-1 is less dominant, winning only 4, while SGPDP wins 5, and PartPSP-2 wins the remaining 3. As a result, we conclude that PartPSP demonstrates outstanding capability for decentralized optimization under privacy protection.

We also need to verify the theoretical results regarding the impact of the partial communication mechanism on PartPSP’s performance. According to our result in Theorem 2, partial communication has a significant influence on PartPSP's theoretical optimization performance. In theory, PartPSP can reduce the dimension $d_s$ of shared parameters through partial communication to mitigate the performance degradation caused by the introduced differential privacy, thereby better addressing the trade-off between privacy protection and optimization performance. Comparing the results of PartPSP-1, PartPSP-2, and SGPDP (a special case of PartPSP where all parameters are shared) in Table \ref{ACCLOSS}, we observe that PartPSP-1, with the smallest $d_s$, achieves the best performance in almost all privacy-preserving experiments. Specifically, comparing all privacy-preserving experiments, PartPSP-1 achieves a 42.11\% relative improvement in average accuracy over PartPSP-2 and a 61.08\% relative improvement over SGPDP. Furthermore, PartPSP-2 also shows a 13.35\% relative improvement in average accuracy compared to SGPDP. These experimental results validate our theoretical findings in Theorem 2 regarding the impact of partial communication on PartPSP.

With the results in Table~\ref{ACCLOSS}, we experimentally verify that partial communication is a powerful tool for PartPSP to better balance the trade-off between privacy protection and optimization performance. Under the same privacy budget, PartPSP can reduce the degradation in algorithm performance caused by the introduced differential privacy by lowering the dimension $d_s$ of shared parameters, thereby achieving superior optimization performance compared to other algorithms. In non-private settings, although differential privacy is no longer satisfied, PartPSP still still remains highly competitive compared with the conventional fully communicated algorithms.In summary, PartPSP is an excellent privacy-preserving decentralized optimization algorithm that serves as a compelling example of a downstream application built on the DPPS protocol, and its flexible model parameter sharing strategy has great potential for future research.
\section{Conclusion}
In this work, we proposed DPPS, a differentially private variant of the Perturbed Push-Sum protocol, which addresses the key challenge of sensitivity estimation in decentralized settings through a novel low-cost method requiring only a single scalar exchange per round. We proved that DPPS satisfies differential privacy and validated the effectiveness of its sensitivity estimates. Building on DPPS as a plug-and-play privacy primitive, we further designed PartPSP, a decentralized non-convex optimization algorithm that integrates partial communication to reduce the dimensionality of perturbed parameters, thereby improving optimization performance under the same privacy budget. We theoretically showed and empirically verified that PartPSP not only effectively optimizes non-convex objectives but also mitigates the utility degradation caused by differential privacy.

\bibliographystyle{IEEEtran}
\bibliography{reference}

\begin{IEEEbiography}[{\includegraphics[width=1in,height=1.25in,clip,keepaspectratio]{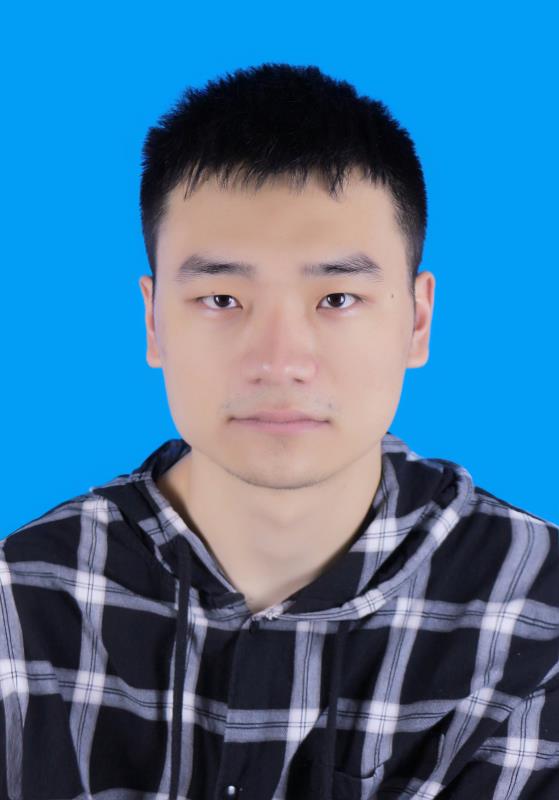}}]{Yiming Zhou}
received his bachelor's degree from the School of Mathematics, Taiyuan University of Technology in July, 2019. He is currently working toward the Ph.D degree from the School of Artificial Intelligence and Data Science, University of Science and Technology of China. His research interests include machine learning, distributed optimization, and network security. He has contributed several papers to high-level journals, including NEURAL NETWORKS and IEEE Transactions on Control of Network Systems.
\end{IEEEbiography}

\begin{IEEEbiography}[{\includegraphics[width=1in,height=1.25in,clip,keepaspectratio]{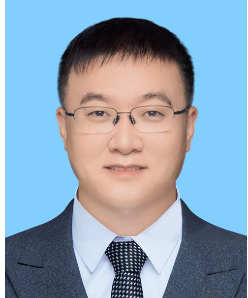}}]{Kaiping Xue}
(Senior Member, IEEE) received the bachelor's degree from the Department of Information Security, University of Science and Technology of China (USTC), in 2003 and received the Ph.D degree from the Department of Electronic Engineering and Information Science (EEIS), USTC, in 2007. From May 2012 to May 2013, he was a postdoctoral researcher with the Department of Electrical and Computer Engineering, University of Florida. Currently, he is a Professor in the School of Cyber Science and Technology, USTC. He is also the director of Network and Information Center, USTC. His research interests include next-generation Internet architecture design, transmission optimization and network security. His work won best paper awards in IEEE MSN 2017 and IEEE HotICN 2019, the Best Paper Honorable Mention in ACM CCS 2022, the Best Paper Runner-Up Award in IEEE MASS 2018, and the best track paper in MSN 2020. He serves on the Editorial Board of several journals, including the IEEE Transactions on Information Forensics and Security (TIFS), the IEEE Transactions on Dependable and Secure Computing (TDSC), the IEEE Transactions on Wireless Communications (TWC), and the IEEE Transactions on Network and Service Management (TNSM). He has served as a Guest Editor for many reputed journals/magazines, and he has also served as a track/symposium chair for some reputed conferences. He is an IET Fellow and an IEEE Senior Member.
\end{IEEEbiography}

\begin{IEEEbiography}[{\includegraphics[width=1in,height=1.25in,clip,keepaspectratio]{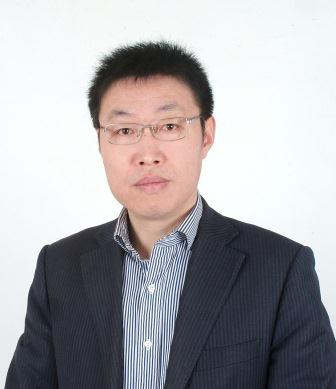}}]{Enhong Chen}
(Fellow, IEEE) received the B.S. degree from Anhui University in 1989, the M.S. degree from Hefei University of Technology in 1992, and the Ph.D. degree in Computer Science from the University of Science and Technology of China (USTC) in 1996. Currently, he is a Professor at the School of Computer Science and Technology, USTC. He is also a CCF Fellow, IEEE Fellow, and ACM Distinguished Member. He is the winner of the National Science Fund for Distinguished Young Scholars (2013) and a member of the Decision Advisory Committee of Shanghai (since June 2018). Additionally, he serves as the Vice Dean of the Faculty of Information and Intelligence at USTC, the Vice Director of the National Engineering Laboratory for Speech and Language Information Processing, the Director of the Anhui Province Key Laboratory of Big Data Analysis and Application, and the Chairman of the Anhui Province Big Data Industry Alliance. His current research interests include data mining and machine learning, particularly in social network analysis and recommendation systems. He has published more than 200 papers in various journals and conferences, including IEEE Transactions, ACM Transactions, and key data mining conferences such as KDD, ICDM, and NeurIPS. He has received several awards, including the Best Application Paper Award at KDD 2008, the Best Research Paper Award at ICDM 2011, the Outstanding Paper Award at FCS 2016, and the Best Student Paper Award at KDD 2024.
\end{IEEEbiography}

\clearpage
\onecolumn

\appendices
\section{Proof of Theorem 2}
\subsection{Lemmas for PartPSP}
\noindent
{\bf Lemma 3}.{ \it
Under Definition 1, for the sequences $\left \{ \bar{\textbf{s} }^{(t)}  \right \} _{t\ge 0}$ in PartPSP, the following equation holds: 
\begin{equation*}
\bar{\textbf{s}}^{(t+1)}= \bar{\textbf{s}}^{(t)}+\frac{\gamma_n}{N}\sum_{i=1}^{N}\textbf{n}_i^{(t)}-\frac{\gamma_s}{N}\sum_{i=1}^{N}\nabla _s^c F_i\left ( \textbf{y} ^{(t)}_{i},\textbf{l}^{(t+1)}_{i};\xi_{i}^{(t)} \right ).
\end{equation*}
}

\begin{proof}
\begin{equation}\label{lemma22eq1}
\begin{aligned}
&& \bar{\textbf{s}}^{(t+1)}&=\frac{1}{N}\sum_{i=1}^{N}\textbf{s}_i^{(t+1)}\\
&&&=\frac{1}{N}\sum_{i=1}^{N}\sum_{j=1}^{N}w_{i,j}^{(t)}\textbf{s}_{j,noise}^{(t+\frac{1}{2} )} \\
&&&=\frac{1}{N}\sum_{j=1}^{N}\textbf{s}_{j,noise}^{(t+\frac{1}{2} )}\sum_{i=1}^{N}w_{i,j}^{(t)} \\
&&&=\frac{1}{N}\sum_{j=1}^{N}\textbf{s}_{j,noise}^{(t+\frac{1}{2} )}\\
&&&=\frac{1}{N}\sum_{i=1}^{N}\textbf{s}_i^{(t)}+\frac{\gamma_n}{N}\sum_{i=1}^{N}\textbf{n}_i^{(t)}-\frac{\gamma_s}{N}\sum_{i=1}^{N}\nabla _s^c F_i\left ( \textbf{y} ^{(t)}_{i},\textbf{l}^{(t+1)}_{i};\xi_{i}^{(t)} \right ).
\end{aligned}
\end{equation}
\end{proof}

The following lemma bounds the changes in the objective function resulting from each node’s sequential update of local and shared parameters in every iteration of PartPSP.

\noindent
{\bf Lemma 4}.{ \it
Suppose that Assumptions 1, 2, and 3 hold and that the weight matrix satisfies Definition 1. Consider the sequences $\left \{ \bar{\textbf{s} }^{(t)}  \right \} _{t\ge 0}$ and $\left \{ \textbf{L} ^{(t)}  \right \} _{t\ge 0}$. For the update to the set of local parameters, we have the following upper bound.
\begin{equation*}
\begin{aligned}
&&&\mathbb{E}\left [F\left ( \bar{\textbf{s} }^{(t)},\textbf{L}^{(t+1)} \right )-F\left ( \bar{\textbf{s} }^{(t)},\textbf{L}^{(t)} \right )\right ]\\
&&&\le \frac{L_{ls}^2}{2N}\sum_{i=1}^{N}\mathbb{E}\left[\left \|\textbf{y}^{(t)}_{i} -\bar{\textbf{s} }^{(t)}\right \|_2^2\right] +(1+L_l)\gamma_l^2\sigma_l^2\\
&&&\quad +\left(\gamma_l^2+L_l\gamma_l^2-\gamma_l\right)\Delta^{(t)}_{\textbf{l} }.
\end{aligned}
\end{equation*}
For the update to the shared parameters, the following upper bound is established.
\begin{equation*}
\begin{aligned}
&&&\mathbb{E}\left [F\left ( \bar{\textbf{s} }^{(t+1)},\textbf{L}^{(t+1)} \right )-F\left ( \bar{\textbf{s} }^{(t)},\textbf{L}^{(t+1)} \right ) \right ]\\
&&&\le \left ( \frac{\gamma_n}{2}+4L_s\gamma_s^2-\frac{\gamma_s}{2} \right ) \Delta^{(t)}_{\bar{\textbf{s} } }+ 4L_s\gamma_s^2 \left ( \sigma_s^2+\sigma_g^2 \right )  \\
&&&\quad +\left ( \frac{\gamma_n}{2N} + \frac{L_s\gamma_s^2}{N}  \right )\sum_{i=1}^N\mathbb{E}\left[ \left \| \textbf{n }_i^{(t)}\right \| _2^2\right]\\
&&&\quad +\left ( \frac{\gamma_sL_{sc}^2}{2}+4L_sL_{sc}^2\gamma_s^2 \right ) \frac{1}{N}\sum_{i=1}^{N}\mathbb{E}\left[\left \|\textbf{y} ^{(t)}_{i}-\bar{\textbf{s} }^{(t)}  \right \| _2^2\right].
\end{aligned}
\end{equation*}
}
\begin{proof}
According to the definition of $F$ in Problem 1, for the update to the set of local parameters, we have:
\begin{equation}\label{theorem23eq2}
\begin{aligned}
&&&\mathbb{E}\left [F\left ( \bar{\textbf{s} }^{(t)},\textbf{L}^{(t+1)} \right )-F\left ( \bar{\textbf{s} }^{(t)},\textbf{L}^{(t)} \right )\right ]\\
&&&= \frac{1}{N}\sum_{i=1}^{N}\mathbb{E} \left [ F_i\left ( \bar{\textbf{s} }^{(t)},\textbf{l}^{(t+1)}_{i} \right )-F_i\left ( \bar{\textbf{s} }^{(t)},\textbf{l}^{(t)}_{i} \right ) \right ]   \\
&&&\le \frac{1}{N}\sum_{i=1}^{N}\mathbb{E} \left [ \left \langle \nabla _l F_i\left ( \bar{\textbf{s} }^{(t)},\textbf{l}^{(t)}_{i} \right ) ,\textbf{l}^{(t+1)}_{i}-\textbf{l}^{(t)}_{i} \right \rangle +\frac{L_l}{2}\left \| \textbf{l}^{(t+1)}_{i}-\textbf{l}^{(t)}_{i} \right \|_2^2   \right ]\\
&&&\le \frac{1}{N}\sum_{i=1}^{N}\underbrace{\mathbb{E} \left \langle \nabla _l F_i\left ( \bar{\textbf{s} }^{(t)},\textbf{l}^{(t)}_{i} \right ) ,\textbf{l}^{(t+1)}_{i}-\textbf{l}^{(t)}_{i} \right \rangle}_{(a1)} + \frac{L_l}{2N}\sum_{i=1}^{N}\underbrace{\mathbb{E}\left \| \textbf{l}^{(t+1)}_{i}-\textbf{l}^{(t)}_{i} \right \|_2^2}_{(a2)},
\end{aligned}
\end{equation}

\noindent
For (a1) in Eq.(\ref{theorem23eq2}), we have
\begin{equation}\label{theorem23eq3}
\begin{aligned}
&&(a1)&=\mathbb{E} \left \langle \nabla _l F_i\left ( \bar{\textbf{s} }^{(t)},\textbf{l}^{(t)}_{i} \right )-\nabla _l F_i\left ( \textbf{y} ^{(t)}_{i},\textbf{l}^{(t)}_{i} \right )+\nabla _l F_i\left ( \textbf{y} ^{(t)}_{i},\textbf{l}^{(t)}_{i} \right ) ,\textbf{l}^{(t+1)}_{i}-\textbf{l}^{(t)}_{i} \right \rangle \\
&&& =\mathbb{E} \left \langle \nabla _l F_i\left ( \bar{\textbf{s} }^{(t)},\textbf{l}^{(t)}_{i} \right )-\nabla _l F_i\left ( \textbf{y} ^{(t)}_{i},\textbf{l}^{(t)}_{i} \right ),\textbf{l}^{(t+1)}_{i}-\textbf{l}^{(t)}_{i} \right \rangle \\
&&&\quad +\mathbb{E} \left \langle \nabla _l F_i\left ( \textbf{y} ^{(t)}_{i},\textbf{l}^{(t)}_{i} \right ),-\gamma _{l}\nabla _l F_i\left ( \textbf{y} ^{(t)}_{i},\textbf{l}^{(t)}_{i};\xi_{i}^{(t)} \right ) \right \rangle\\
&&&\le \frac{1}{2} \mathbb{E}\left \| \nabla _l F_i\left ( \bar{\textbf{s} }^{(t)},\textbf{l}^{(t)}_{i} \right )-\nabla _l F_i\left ( \textbf{y} ^{(t)}_{i},\textbf{l}^{(t)}_{i} \right ) \right \| _2^2+\frac{1}{2} \mathbb{E}\left \| \textbf{l}^{(t+1)}_{i}-\textbf{l}^{(t)}_{i} \right \| _2^2-\gamma_l\mathbb{E}\left \| \nabla _l F_i\left ( \textbf{y} ^{(t)}_{i},\textbf{l}^{(t)}_{i} \right ) \right \| _2^2\\
&&&\le \frac{L_{ls}^2}{2}\mathbb{E}\left \|\bar{\textbf{s} }^{(t)}- \textbf{y} ^{(t)}_{i}\right \|_2^2+\frac{1}{2} \mathbb{E}\left \| \textbf{l}^{(t+1)}_{i}-\textbf{l}^{(t)}_{i} \right \| _2^2-\gamma_l\mathbb{E}\left \| \nabla _l F_i\left ( \textbf{y} ^{(t)}_{i},\textbf{l}^{(t)}_{i} \right ) \right \| _2^2.
\end{aligned}
\end{equation}

\noindent
For (a2) in Eq.(\ref{theorem23eq2}), we have
\begin{equation}\label{theorem23eq4}
\begin{aligned}
&&(a2)&=\gamma_l^2\mathbb{E} \left \| \nabla _l F_i\left ( \textbf{y} ^{(t)}_{i},\textbf{l}^{(t)}_{i};\xi_{i}^{(t)} \right )-\nabla _l F_i\left ( \textbf{y} ^{(t)}_{i},\textbf{l}^{(t)}_{i} \right ) +\nabla _l F_i\left ( \textbf{y} ^{(t)}_{i},\textbf{l}^{(t)}_{i}\right )\right \| _2^2\\
&&&\le 2\gamma_l^2\mathbb{E} \left \| \nabla _l F_i\left ( \textbf{y} ^{(t)}_{i},\textbf{l}^{(t)}_{i};\xi_{i}^{(t)} \right )-\nabla _l F_i\left ( \textbf{y} ^{(t)}_{i},\textbf{l}^{(t)}_{i} \right ) \right \| _2^2+2\gamma_l^2\mathbb{E} \left \| \nabla _l F_i\left ( \textbf{y} ^{(t)}_{i},\textbf{l}^{(t)}_{i}\right ) \right \| _2^2\\
&&&\le 2\gamma_l^2\sigma_l^2 + 2\gamma_l^2\mathbb{E} \left \| \nabla _l F_i\left ( \textbf{y} ^{(t)}_{i},\textbf{l}^{(t)}_{i}\right ) \right \| _2^2.
\end{aligned}
\end{equation}

\noindent
Back to Eq.(\ref{theorem23eq2}), we have
\begin{equation}\label{theorem23eq5}
\begin{aligned}
&&&\mathbb{E}\left [F\left ( \bar{\textbf{s} }^{(t)},\textbf{L}^{(t+1)} \right )-F\left ( \bar{\textbf{s} }^{(t)},\textbf{L}^{(t)} \right )\right ]\\
&&&\le \frac{L_{ls}^2}{2N}\sum_{i=1}^{N}\mathbb{E}\left \|\bar{\textbf{s} }^{(t)}- \textbf{y} ^{(t)}_{i}\right \|_2^2+(1+L_l)\gamma_l^2\sigma_l^2+\frac{(1+L_l)\gamma_l^2-\gamma_l}{N}\sum_{i=1}^{N}\mathbb{E} \left \| \nabla _l F_i\left ( \textbf{y} ^{(t)}_{i},\textbf{l}^{(t)}_{i}\right ) \right \| _2^2\\
&&&\le \frac{L_{ls}^2}{2N}\sum_{i=1}^{N}\mathbb{E}\left[\left \|\textbf{y}^{(t)}_{i} -\bar{\textbf{s} }^{(t)}\right \|_2^2\right] +(1+L_l)\gamma_l^2\sigma_l^2 +\left(\gamma_l^2+L_l\gamma_l^2-\gamma_l\right)\Delta^{(t)}_{\textbf{l} }.
\end{aligned}
\end{equation}

\noindent
According to the smoothness in Assumption 2, for the update to the shared parameters, we have:
\begin{equation}\label{theorem23eq6}
\begin{aligned}
&&&\mathbb{E}\left [F\left ( \bar{\textbf{s} }^{(t+1)},\textbf{L}^{(t+1)} \right )-F\left ( \bar{\textbf{s} }^{(t)},\textbf{L}^{(t+1)} \right ) \right ]\\
&&& \le \mathbb{E} \left [ \left \langle \nabla _s F\left ( \bar{\textbf{s} }^{(t)},\textbf{L}^{(t+1)} \right ) ,\bar{\textbf{s} }^{(t+1)}-\bar{\textbf{s} }^{(t)} \right \rangle +\frac{L_s}{2}\left \| \bar{\textbf{s} }^{(t+1)}-\bar{\textbf{s} }^{(t)} \right \|_2^2   \right ]\\
&&& \le \underbrace{\mathbb{E} \left [ \frac{\gamma_n}{N}\sum_{i=1}^{N}\left \langle \nabla _s F\left ( \bar{\textbf{s} }^{(t)},\textbf{L}^{(t+1)} \right ) ,\textbf{n}_i^{(t)} \right \rangle\right]}_{(b1)}\\
&&&\quad \underbrace{-\mathbb{E} \left [\frac{\gamma_s}{N}\sum_{i=1}^{N}\left \langle \nabla _s F\left ( \bar{\textbf{s} }^{(t)},\textbf{L}^{(t+1)} \right ) ,\nabla _s^c F_i\left ( \textbf{y}_i^{(t)},\textbf{l}_i^{(t+1)};\xi_i^{(t)} \right ) \right \rangle\right]}_{(b2)}\\
&&&\quad \underbrace{+\mathbb{E} \left [\frac{L_s}{2}\left \| \bar{\textbf{s} }^{(t+1)}-\bar{\textbf{s} }^{(t)} \right \|_2^2 \right]}_{(b3)}.
\end{aligned}
\end{equation}

\noindent
For (b1) in Eq.(\ref{theorem23eq6}), we have
\begin{equation}\label{theorem23eq7}
\begin{aligned}
&&(b1)&\le \frac{\gamma_n}{2} \mathbb{E}\left \| \nabla _s F\left ( \bar{\textbf{s} }^{(t)},\textbf{L}^{(t+1)} \right ) \right \| _2^2+ \frac{\gamma_n}{2N}\sum_{i=1}^{N}\mathbb{E}\left \|  \textbf{n}_i^{(t)}\right \|_2^2 .
\end{aligned}
\end{equation}

\noindent
For (b2) in Eq.(\ref{theorem23eq6}), we have
\begin{equation}\label{theorem23eq8}
\begin{aligned}
&&(b2)& = -\mathbb{E} \left [\frac{\gamma_s}{N}\sum_{i=1}^{N}\left \langle \nabla _s F\left ( \bar{\textbf{s} }^{(t)},\textbf{L}^{(t+1)} \right ) ,\nabla _s^c F_i\left ( \textbf{y}_i^{(t)},\textbf{l}_i^{(t+1)} \right )- \nabla _s F_i\left ( \bar{\textbf{s} }^{(t)},\textbf{l}_i^{(t+1)} \right )+\nabla _s F_i\left ( \bar{\textbf{s} }^{(t)},\textbf{l}_i^{(t+1)} \right )\right \rangle\right]\\
&&&=-\gamma_s\mathbb{E} \left \| \nabla _s F\left ( \bar{\textbf{s} }^{(t)},\textbf{L}^{(t+1)} \right ) \right \| _2^2 \\
&&&\quad + \mathbb{E} \left [\frac{\gamma_s}{N}\sum_{i=1}^{N}\left \langle \nabla _s F\left ( \bar{\textbf{s} }^{(t)},\textbf{L}^{(t+1)} \right ) ,\nabla _s F_i\left ( \bar{\textbf{s} }^{(t)},\textbf{l}_i^{(t+1)} \right )-\nabla _s^c F_i\left ( \textbf{y}_i^{(t)},\textbf{l}_i^{(t+1)} \right )\right \rangle\right]\\
&&&\le -\gamma_s\mathbb{E} \left \| \nabla _s F\left ( \bar{\textbf{s} }^{(t)},\textbf{L}^{(t+1)} \right ) \right \| _2^2 \\
&&&\quad + \frac{\gamma_s}{2N}\sum_{i=1}^N\mathbb{E} \left [ \left \| \nabla _s F\left ( \bar{\textbf{s} }^{(t)},\textbf{L}^{(t+1)} \right ) \right \|_2^2 +\left \| \nabla _s F_i\left ( \bar{\textbf{s} }^{(t)},\textbf{l}_i^{(t+1)} \right )-\nabla _s^c F_i\left ( \textbf{y}_i^{(t)},\textbf{l}_i^{(t+1)} \right ) \right \|_2^2\right ]\\
&&&\le -\frac{\gamma_s}{2}\mathbb{E} \left \| \nabla _s F\left ( \bar{\textbf{s} }^{(t)},\textbf{L}^{(t+1)} \right ) \right \| _2^2+\frac{\gamma_sL_{sc}^2}{2N}\sum_{i=1}^N\mathbb{E}\left \|\bar{\textbf{s} }^{(t)}- \textbf{y}_i^{(t)} \right \| _2^2.
\end{aligned}
\end{equation}

\noindent
For (b3) in Eq.(\ref{theorem23eq6}), according to Lemma 3, we have
\begin{equation}\label{theorem23eq9}
\begin{aligned}
&&(b3)& = \frac{L_s}{2}\mathbb{E} \left [\left \| \frac{\gamma_s}{N}\sum_{i=1}^{N}\textbf{n}_i^{(t)}-\frac{\gamma_s}{N}\sum_{i=1}^{N}\nabla _s^c F_i\left ( \textbf{y} ^{(t)}_{i},\textbf{l}^{(t+1)}_{i};\xi_{i}^{(t)} \right ) \right \|_2^2 \right]\\
&&&\le \frac{L_s\gamma_s^2}{N^2}\mathbb{E}\left \|\sum_{i=1}^{N}\textbf{n}_i^{(t)} \right \|_2^2 + \frac{L_s\gamma_s^2}{N^2}\mathbb{E}\left \|\sum_{i=1}^{N}\nabla _s^c F_i\left ( \textbf{y} ^{(t)}_{i},\textbf{l}^{(t+1)}_{i};\xi_{i}^{(t)} \right ) \right \|_2^2\\
&&&\le \frac{L_s\gamma_s^2}{N}\sum_{i=1}^{N}\mathbb{E}\left \|\textbf{n}_i^{(t)} \right \|_2^2 +\frac{L_s\gamma_s^2}{N}\sum_{i=1}^{N}\mathbb{E}\left \|\nabla _s^c F_i\left ( \textbf{y} ^{(t)}_{i},\textbf{l}^{(t+1)}_{i};\xi_{i}^{(t)} \right ) \right \|_2^2\\
&&&\le \frac{L_s\gamma_s^2}{N}\sum_{i=1}^{N}\mathbb{E}\left \|\textbf{n}_i^{(t)} \right \|_2^2+\frac{4L_s\gamma_s^2}{N}\sum_{i=1}^{N}\mathbb{E}\left \|\nabla _s^c F_i\left ( \textbf{y} ^{(t)}_{i},\textbf{l}^{(t+1)}_{i};\xi_{i}^{(t)} \right )-\nabla _s^c F_i\left ( \textbf{y} ^{(t)}_{i},\textbf{l}^{(t+1)}_{i} \right )\right \|_2^2\\
&&&\quad +\frac{4L_s\gamma_s^2}{N}\sum_{i=1}^{N}\mathbb{E}\left \|\nabla _s^c F_i\left ( \textbf{y} ^{(t)}_{i},\textbf{l}^{(t+1)}_{i} \right )- \nabla _s^c F\left ( \textbf{y} ^{(t)}_{i},\textbf{L}^{(t+1)} \right )\right \|_2^2\\
&&&\quad +\frac{4L_s\gamma_s^2}{N}\sum_{i=1}^{N}\mathbb{E}\left\|\nabla _s^c F\left ( \textbf{y} ^{(t)}_{i},\textbf{L}^{(t+1)} \right )- \nabla _s F\left ( \bar{\textbf{s} }^{(t)},\textbf{L}^{(t+1)} \right )\right \|_2^2\\
&&&\quad +\frac{4L_s\gamma_s^2}{N}\sum_{i=1}^{N}\mathbb{E}\left\|\nabla _s F\left ( \bar{\textbf{s} }^{(t)},\textbf{L}^{(t+1)} \right )\right \|_2^2\\
&&&\le \frac{L_s\gamma_s^2}{N}\sum_{i=1}^{N}\mathbb{E}\left \|\textbf{n}_i^{(t)} \right \|_2^2 +4L_s \gamma_s^2 \sigma_s^2+4L_s \gamma_s^2 \sigma_g^2+\frac{4L_sL_{sc}^2\gamma_s^2}{N}\sum_{i=1}^{N}\mathbb{E}\left \|\textbf{y} ^{(t)}_{i}-\bar{\textbf{s} }^{(t)}  \right \| _2^2 \\
&&&\quad +4L_s\gamma_s^2\mathbb{E}\left\|\nabla _s F\left ( \bar{\textbf{s} }^{(t)},\textbf{L}^{(t+1)} \right )\right \|_2^2.
\end{aligned}
\end{equation}

\noindent
Back to Eq.(\ref{theorem23eq6}), we have
\begin{equation}\label{theorem23eq10}
\begin{aligned}
&&&\mathbb{E}\left [F\left ( \bar{\textbf{s} }^{(t+1)},\textbf{L}^{(t+1)} \right )-F\left ( \bar{\textbf{s} }^{(t)},\textbf{L}^{(t+1)} \right ) \right ]\\
&&& \le \left ( \frac{\gamma_n}{2}+4L_s\gamma_s^2-\frac{\gamma_s}{2} \right ) \Delta^{(t)}_{\bar{\textbf{s} } }+\left ( \frac{\gamma_n}{2N} + \frac{L_s\gamma_s^2}{N}  \right )\sum_{i=1}^N\mathbb{E}\left[ \left \| \textbf{n }_i^{(t)}\right \| _2^2\right]\\
&&&\quad +4L_s\gamma_s^2 \left ( \sigma_s^2+\sigma_g^2 \right )  +\left ( \frac{\gamma_sL_{sc}^2}{2}+4L_sL_{sc}^2\gamma_s^2 \right ) \frac{1}{N}\sum_{i=1}^{N}\mathbb{E}\left \|\textbf{y} ^{(t)}_{i}-\bar{\textbf{s} }^{(t)}  \right \| _2^2.\\
\end{aligned}
\end{equation}
\end{proof}

The following lemma provides an upper bound on the variance of the noise vector in PartPSP under Assumption 4.

\noindent
{\bf Lemma 5}.{ \it
Assuming Assumption 4 holds, consider the noise vector $\mathbf{n}_i^{(t)}$ for all $i \in [N]$ at iteration $t \geq 0$. The following bound holds:
\begin{equation*}
\mathbb{E}\left[ \left \| \mathbf{n}_i^{(t)} \right \|_2^2 \right] \leq \frac{2d_s S^2}{b^2}.
\end{equation*}
}

\begin{proof}
\begin{equation}
\mathbb{E}\left [\left \| \textbf{n }\right \| _2^2\right ] = \mathbb{E}\left [ \sum_{i=1}^{d_s} n_i^2\right ] \overset{(a)}{=}\sum_{i=1}^{d_s}\mathbb{E}\left [\left ( n_i - \mathbb{E}n_i \right ) ^2\right ]=\sum_{i=1}^{d_s}Var\left ( Lap(0,\frac{S}{b} ) \right ) =\frac{2d_sS^2}{b^2},
\end{equation}
where (a) follows by $\mathbb{E}\left [n_i\right ]=0$.
\end{proof}

The following lemma provides an upper bound on the distance between $\textbf{y}$ and $\bar{\textbf{s}}$ across the network.

\noindent
{\bf Lemma 6}.{ \it
Suppose that Assumptions 1, 2, and 3 hold and that the weight matrix satisfies Definition 1. Consider the sequences $\left \{ \textbf{y}_{i}^{(t)}  \right \} _{t\ge 0}$ and $\left \{ \bar{\textbf{s} } ^{(t)}  \right \} _{t\ge 0}$. Under the condition $\gamma_s \le \frac{1-q}{2\sqrt{15} CL_{sc}} $, we have the following upper bound.
\begin{equation*}
\begin{aligned}
&&&\frac{1}{N}\sum_{i=1}^N\sum_{t=0}^{T-1}\mathbb{E}\left[\left \|\textbf{y}^{(t)}_i- \bar{\textbf{s} }^{(t)} \right \|_2^2\right]\\
&&&\le \frac{60d_s\gamma_n^2 C^2S^2T}{b^2(1-q)^2} + \frac{30\gamma_s^2 C^2T\left ( \sigma_s^2 +\sigma_g^2\right ) }{(1-q)^2}  \\
&&&\quad +\frac{10C^2\sum_{i=1}^N\left\|\textbf{s}^{(0)}_i\right\|_2^2}{N(1-q^2)}+\frac{60\gamma_s^2 C^2}{(1-q)^2}\sum_{t=0}^{T-1}\Delta^{(t)}_{\bar{\textbf{s} } }.
\end{aligned}
\end{equation*}
where $C$ is a constant and $q\in (0,1)$.
}

\begin{proof}
From Lemma 3 in Assran \textit{et al.} \cite{SGP}, we have 

\begin{equation}\label{theorem22eq1}
\begin{aligned}
&&&\left \|\textbf{y}^{(t)}_i- \bar{\textbf{s} }^{(t)} \right \|_2\\
&&\ &\le C q^{t}\left \|\textbf{s}^{(0)}_i\right \|_2 + C \sum_{k=0}^t q^{t-k}\left \|\gamma_n\textbf{n}^{(k)}_i-\gamma_s\nabla _s^c F_i \left ( \textbf{y}_{i}^{(k)},\textbf{l}_{i}^{(k+1)};\xi ^{(k)}_{i} \right )  \right \|_2\\
&&\ &\le C q^{t}\left \|\textbf{s}^{(0)}_i\right \|_2 + \gamma_n C \sum_{k=0}^t q^{t-k}\left \|\textbf{n}^{(k)}_i\right \|_2 + \gamma_s C \sum_{k=0}^t q^{t-k}\left \| \nabla _s^c F_i \left ( \textbf{y}_{i}^{(k)},\textbf{l}_{i}^{(k+1)};\xi ^{(k)}_{i} \right ) - \nabla _s^c F_i \left ( \textbf{y}_{i}^{(k)},\textbf{l}_{i}^{(k+1)} \right ) \right \|_2\\
&&&\quad + \gamma_s C \sum_{k=0}^t q^{t-k}\left \| \nabla _s^c F_i \left ( \textbf{y}_{i}^{(k)},\textbf{l}_{i}^{(k+1)} \right ) - \nabla _s^c F \left ( \textbf{y}_{i}^{(k)},\textbf{L}^{(k+1)} \right ) \right \|_2 + \gamma_s C \sum_{k=0}^t q^{t-k}\left \| \nabla _s^c F \left ( \textbf{y}_{i}^{(k)},\textbf{L}^{(k+1)} \right ) \right \|_2.
\end{aligned}
\end{equation}

\noindent
Square both sides of Eq.(\ref{theorem22eq1}) and take the mathematical expectation. Consider the Cauchy-Schwarz inequality $\mathbb{E}  (\textbf{a} + \textbf{b} + \textbf{c}+ \textbf{d}+ \textbf{e} )^2 \le 5\mathbb{E} \textbf{a} ^2 +5\mathbb{E} \textbf{b}^2 +5\mathbb{E} \textbf{c}^2 +5\mathbb{E} \textbf{d} ^2 +5\mathbb{E} \textbf{e} ^2$, we let

\begin{flalign*}
&\textbf{a} = C q^{t}\left\|\textbf{s}^{(0)}_i\right\|_2,& \\
&\textbf{b}=\sum_{k=0}^t q^{t-k}\left\|\textbf{n}^{(k)}_i\right\|_2,& \\
&\textbf{c}=\sum_{k=0}^t q^{t-k}\left\| \nabla _s^c F_i \left ( \textbf{y}_{i}^{(k)},\textbf{l}_{i}^{(k+1)};\xi ^{(k)}_{i} \right ) - \nabla _s^c F_i \left ( \textbf{y}_{i}^{(k)},\textbf{l}_{i}^{(k+1)} \right ) \right\|_2,& \\
&\textbf{d}=\sum_{k=0}^t q^{t-k}\left\| \nabla _s^c F_i \left ( \textbf{y}_{i}^{(k)},\textbf{l}_{i}^{(k+1)} \right ) - \nabla _s^c F \left ( \textbf{y}_{i}^{(k)},\textbf{L}^{(k+1)} \right ) \right\|_2,& \\
&\textbf{e}=\sum_{k=0}^t q^{t-k}\left\| \nabla _s^c F \left ( \textbf{y}_{i}^{(k)},\textbf{L}^{(k+1)} \right ) \right\|_2.&
\end{flalign*}

\noindent
For $\textbf{a}$, we have
\begin{equation}\label{theorem22eq2}
\mathbb{E} \textbf{a}^2 = C^2 q^{2t} \left\|\textbf{s}^{(0)}_i\right\|_2^2.
\end{equation}

\noindent
For $\textbf{b}$, we have
\begin{equation}\label{theorem22eq3}
\begin{aligned}
&&\mathbb{E} \textbf{b}^2 &= \sum_{k=0}^t q^{2(t-k)}\mathbb{E} \left\|\textbf{n}^{(k)}_i\right\|_2^2 +2\sum_{j=0}^{t}\sum_{k=j+1}^{t} q^{2t-j-k}\left \langle \mathbb{E} \left\|\textbf{n}^{(j)}_i\right\|_2,\mathbb{E} \left\|\textbf{n}^{(k)}_i\right\|_2 \right \rangle \\
&&\ &\le \sum_{k=0}^t q^{2(t-k)}\mathbb{E} \left\|\textbf{n}^{(k)}_i\right\|_2^2 +\sum_{j=0}^{t}\sum_{k=j+1}^{t} q^{2t-j-k}\left (\mathbb{E} \left\|\textbf{n}^{(j)}_i\right\|_2^2+\mathbb{E} \left\|\textbf{n}^{(k)}_i\right\|_2^2 \right )\\
&&\ &\le \sum_{k=0}^t q^{2(t-k)}\mathbb{E} \left\|\textbf{n}^{(k)}_i\right\|_2^2 +2\sum_{j=0}^{t}\sum_{k=0}^{t} q^{2t-j-k}\mathbb{E} \left\|\textbf{n}^{(j)}_i\right\|_2^2\\
&&\ &\le \sum_{k=0}^t q^{2(t-k)}\mathbb{E} \left\|\textbf{n}^{(k)}_i\right\|_2^2 +2\sum_{j=0}^{t} q^{t-j}\mathbb{E} \left\|\textbf{n}^{(j)}_i\right\|_2^2\sum_{k=0}^{t} q^{t-k}\\
&&\ &\le \sum_{k=0}^t q^{2(t-k)}\mathbb{E} \left\|\textbf{n}^{(k)}_i\right\|_2^2 +\frac{2}{1-q}\sum_{k=0}^{t} q^{t-k}\mathbb{E} \left\|\textbf{n}^{(k)}_i\right\|_2^2\\
&&\ &\le \frac{3}{1-q}\sum_{k=0}^{t} q^{t-k}\mathbb{E} \left\|\textbf{n}^{(k)}_i\right\|_2^2\\
&&\ &\le \frac{6d_sS^2}{b^2(1-q)^2}.
\end{aligned}
\end{equation}

\noindent
In the same way, we have
\begin{equation}\label{theorem22eq4}
\mathbb{E} \textbf{c}^2 \le \frac{3\sigma_s^2}{(1-q)^2},
\end{equation}
\begin{equation}\label{theorem22eq5}
\mathbb{E} \textbf{d}^2 \le \frac{3\sigma_g^2}{(1-q)^2}.
\end{equation}

\noindent
For $\textbf{e}$, similar with $\textbf{b}$, we have
\begin{equation}\label{theorem22eq6}
\begin{aligned}
&&\mathbb{E} \textbf{e}^2 &\le \frac{3}{1-q}\sum_{k=0}^t q^{t-k}\mathbb{E}\left [\left\| \nabla _s^c F \left ( \textbf{y}_{i}^{(k)},\textbf{L}^{(k+1)} \right ) \right\|_2^2\right ] \\
&&& \le \frac{3}{1-q}\sum_{k=0}^t q^{t-k}\mathbb{E}\left [\left\| \nabla _s^c F \left ( \textbf{y}_{i}^{(k)},\textbf{L}^{(k+1)} \right ) - \nabla _s F \left ( \bar{\textbf{s} }^{(k)} ,\textbf{L}^{(k+1)} \right ) + \nabla _s F \left ( \bar{\textbf{s} }^{(k)} ,\textbf{L}^{(k+1)} \right )\right\|_2^2\right ] \\
&&& \le \frac{6}{1-q}\sum_{k=0}^t q^{t-k}\mathbb{E}\left [\left\| \nabla _s^c F \left ( \textbf{y}_{i}^{(k)},\textbf{L}^{(k+1)} \right ) - \nabla _s F \left ( \bar{\textbf{s} }^{(k)} ,\textbf{L}^{(k+1)} \right )\right\|_2^2 +\left\|  \nabla _s F \left ( \bar{\textbf{s} }^{(k)} ,\textbf{L}^{(k+1)} \right )\right\|_2^2\right ] \\
&&& \le \frac{6L_{sc}^2}{1-q}\sum_{k=0}^t q^{t-k}\mathbb{E}\left [\left\| \textbf{y}_{i}^{(k)} - \bar{\textbf{s} }^{(k)}\right\|_2^2 \right ]+\frac{6}{1-q}\sum_{k=0}^t q^{t-k}\mathbb{E}\left [\left\| \nabla _s F \left ( \bar{\textbf{s} }^{(k)},\textbf{L}^{(k+1)} \right ) \right\|_2^2\right ].
\end{aligned}
\end{equation}

\noindent
Back to Eq.(\ref{theorem22eq1}), we have
\begin{equation}\label{theorem22eq7}
\begin{aligned}
&&\mathbb{E}\left \|\textbf{y}^{(t)}_i- \bar{\textbf{s} }^{(t)} \right \|_2^2& \le 5C^2 q^{2t} \left\|\textbf{s}^{(0)}_i\right\|_2^2 +\frac{30d_s\gamma_n^2 C^2S^2}{b^2(1-q)^2} + \frac{15\gamma_s^2 C^2\sigma_s^2}{(1-q)^2} + \frac{15\gamma_s^2 C^2\sigma_g^2}{(1-q)^2} \\
&&&+ \frac{30\gamma_s^2 C^2L_{sc}^2}{1-q}\sum_{k=0}^t q^{t-k}\mathbb{E}\left [\left\| \textbf{y}_{i}^{(k)} - \bar{\textbf{s} }^{(k)}\right\|_2^2 \right ]+\frac{30\gamma_s^2 C^2}{1-q}\sum_{k=0}^t q^{t-k}\mathbb{E}\left [\left\| \nabla _s F \left ( \bar{\textbf{s} }^{(k)},\textbf{L}^{(k+1)} \right ) \right\|_2^2\right ].
\end{aligned}
\end{equation}

\noindent
Sum up Eq.(\ref{theorem22eq7}) over $t\in \left \{ 0, 1,..., T-1 \right \}$, we have
\begin{equation}\label{theorem22eq8}
\begin{aligned}
&&&\sum_{t=0}^{T-1}\mathbb{E}\left \|\textbf{y}^{(t)}_i- \bar{\textbf{s} }^{(t)} \right \|_2^2\\
&&&\le \frac{5C^2\left\|\textbf{s}^{(0)}_i\right\|_2^2}{1-q^2} +\frac{30d_s\gamma_n^2 C^2S^2T}{b^2(1-q)^2} + \frac{15\gamma_s^2 C^2\sigma_s^2T}{(1-q)^2} + \frac{15\gamma_s^2 C^2\sigma_g^2T}{(1-q)^2}\\
&&&+ \frac{30\gamma_s^2 C^2L_{sc}^2}{1-q}\underbrace{\sum_{t=0}^{T-1}\sum_{k=0}^t q^{t-k}\mathbb{E}\left [\left\| \textbf{y}_{i}^{(k)} - \bar{\textbf{s} }^{(k)}\right\|_2^2 \right ]}_{(a)}+\frac{30\gamma_s^2 C^2}{1-q}\underbrace{\sum_{t=0}^{T-1}\sum_{k=0}^t q^{t-k}\mathbb{E}\left [\left\| \nabla _s F \left ( \bar{\textbf{s} }^{(k)},\textbf{L}^{(k+1)} \right ) \right\|_2^2\right ]}_{(b)}.\\
\end{aligned}
\end{equation}

\noindent
For (a) in Eq.(\ref{theorem22eq8}), we have
\begin{equation}\label{theorem22eq9}
\begin{aligned}
&&(a)&=  \sum_{k=0}^{T-1}\sum_{t=k}^{T-1} q^{t-k}\mathbb{E}\left [\left\| \textbf{y}_{i}^{(k)} - \bar{\textbf{s} }^{(k)}\right\|_2^2 \right ]\\
&&& = \sum_{k=0}^{T-1}\mathbb{E}\left [\left\| \textbf{y}_{i}^{(k)} - \bar{\textbf{s} }^{(k)}\right\|_2^2 \right ]\sum_{t=k}^{T-1} q^{t-k}\\
&&& \le \frac{1}{1-q}\sum_{t=0}^{T-1}\mathbb{E}\left [\left\| \textbf{y}_{i}^{(t)} - \bar{\textbf{s} }^{(t)}\right\|_2^2 \right ].
\end{aligned}
\end{equation}

\noindent
In the same way, we have
\begin{equation}\label{theorem22eq10}
(b)\le \frac{1}{1-q}\sum_{t=0}^{T-1}\mathbb{E}\left [\left\| \nabla _s F \left ( \bar{\textbf{s} }^{(k)},\textbf{L}^{(k+1)} \right ) \right\|_2^2\right ].
\end{equation}

\noindent
Back to Eq.(\ref{theorem22eq8}), we have
\begin{equation}\label{theorem22eq11}
\begin{aligned}
&&&\sum_{t=0}^{T-1}\mathbb{E}\left \|\textbf{y}^{(t)}_i- \bar{\textbf{s} }^{(t)} \right \|_2^2\\
&&&\le \frac{5C^2\left\|\textbf{s}^{(0)}_i\right\|_2^2}{1-q^2} +\frac{30d_s\gamma_n^2 C^2S^2T}{b^2(1-q)^2} + \frac{15\gamma_s^2 C^2\sigma_s^2T}{(1-q)^2} + \frac{15\gamma_s^2 C^2\sigma_g^2T}{(1-q)^2}  \\
&&&+ \frac{30\gamma_s^2 C^2L_{sc}^2}{(1-q)^2}\sum_{t=0}^{T-1}\mathbb{E}\left [\left\| \textbf{y}_{i}^{(t)} - \bar{\textbf{s} }^{(t)}\right\|_2^2 \right ]+\frac{30\gamma_s^2 C^2}{(1-q)^2}\sum_{t=0}^{T-1}\mathbb{E}\left [\left\| \nabla _s F \left ( \bar{\textbf{s} }^{(t)},\textbf{L}^{(t+1)} \right ) \right\|_2^2\right ].\\
\end{aligned}
\end{equation}

\noindent
Rearrange Eq.(\ref{theorem22eq11}), we have
\begin{equation}\label{theorem22eq12}
\begin{aligned}
&&&\left ( 1- \frac{30\gamma_s^2 C^2L_{sc}^2}{(1-q)^2}\right ) \sum_{t=0}^{T-1}\mathbb{E}\left \|\textbf{y}^{(t)}_i- \bar{\textbf{s} }^{(t)} \right \|_2^2\\
&&&\le \frac{5C^2\left\|\textbf{s}^{(0)}_i\right\|_2^2}{1-q^2} +\frac{30d_s\gamma_n^2 C^2S^2T}{b^2(1-q)^2} + \frac{15\gamma_s^2 C^2\sigma_s^2T}{(1-q)^2} + \frac{15\gamma_s^2 C^2\sigma_g^2T}{(1-q)^2} \\
&&&\quad +\frac{30\gamma_s^2 C^2}{(1-q)^2}\sum_{t=0}^{T-1}\mathbb{E}\left [\left\| \nabla _s F \left ( \bar{\textbf{s} }^{(t)},\textbf{L}^{(t+1)} \right ) \right\|_2^2\right ].
\end{aligned}
\end{equation}

\noindent
Let $Q = 1- \frac{30\gamma_s^2 C^2L_{sc}^2}{(1-q)^2}$, since $\gamma_s\le\frac{1-q}{2\sqrt{15}CL_{sc}}$, we have $Q\ge \frac{1}{2}$. Divide both side of Eq.(\ref{theorem22eq12}) by $Q$, we have the conclusion

\begin{equation}\label{theorem22eq13}
\begin{aligned}
&&\sum_{t=0}^{T-1}\mathbb{E}\left \|\textbf{y}^{(t)}_i- \bar{\textbf{s} }^{(t)} \right \|_2^2&\le \frac{10C^2\left\|\textbf{s}^{(0)}_i\right\|_2^2}{(1-q^2)} +\frac{60d_s\gamma_n^2 C^2S^2T}{b^2(1-q)^2} + \frac{30\gamma_s^2 C^2\sigma_s^2T}{(1-q)^2} + \frac{30\gamma_s^2 C^2\sigma_g^2T}{(1-q)^2} \\
&&&+\frac{60\gamma_s^2 C^2}{(1-q)^2}\sum_{t=0}^{T-1}\mathbb{E}\left [\left\| \nabla _s F \left ( \bar{\textbf{s} }^{(t)},\textbf{L}^{(t+1)} \right ) \right\|_2^2\right ].
\end{aligned}
\end{equation}

\noindent
Sum up Eq.(\ref{theorem22eq13}) over $i\in \left \{ 1, 2,..., N \right \}$ and divide $N$, we have
\begin{equation}\label{theorem22eq14}
\begin{aligned}
&&\frac{1}{N}\sum_{i=1}^N\sum_{t=0}^{T-1}\mathbb{E}\left \|\textbf{y}^{(t)}_i- \bar{\textbf{s} }^{(t)} \right \|_2^2&\le \frac{10C^2\sum_{i=1}^N\left\|\textbf{s}^{(0)}_i\right\|_2^2}{N(1-q^2)} +\frac{60d_s\gamma_n^2 C^2S^2T}{b^2(1-q)^2} + \frac{30\gamma_s^2 C^2\sigma_s^2T}{(1-q)^2} + \frac{30\gamma_s^2 C^2\sigma_g^2T}{(1-q)^2} \\
&&&+\frac{60\gamma_s^2 C^2}{(1-q)^2}\sum_{t=0}^{T-1}\mathbb{E}\left [\left\| \nabla _s F \left ( \bar{\textbf{s} }^{(t)},\textbf{L}^{(t+1)} \right ) \right\|_2^2\right ]\\
&&&\le \frac{60d_s\gamma_n^2 C^2S^2T}{b^2(1-q)^2} + \frac{30\gamma_s^2 C^2\sigma_s^2T}{(1-q)^2} + \frac{30\gamma_s^2 C^2\sigma_g^2T}{(1-q)^2} \\
&&&\quad+ \frac{10C^2\sum_{i=1}^N\left\|\textbf{s}^{(0)}_i\right\|_2^2}{N(1-q^2)}+\frac{60\gamma_s^2 C^2}{(1-q)^2}\sum_{t=0}^{T-1}\Delta^{(t)}_{\bar{\textbf{s} } }.
\end{aligned}
\end{equation}
\end{proof}

\subsection{Main Proof of Theorem 2}

\begin{proof}
We start with:
\begin{equation}\label{theorem2eq1}
\begin{aligned}
&&&\mathbb{E} \left [ F\left ( \bar{\textbf{s} }^{(t+1)},\textbf{L}^{(t+1)} \right ) -F\left ( \bar{\textbf{s} }^{(t)},\textbf{L}^{(t)} \right ) \right ] \\
&&&=\mathbb{E}\left [F\left ( \bar{\textbf{s} }^{(t)},\textbf{L}^{(t+1)} \right )-F\left ( \bar{\textbf{s} }^{(t)},\textbf{L}^{(t)} \right )\right ] \\
&&&\quad +\mathbb{E}\left [F\left ( \bar{\textbf{s} }^{(t+1)},\textbf{L}^{(t+1)} \right )-F\left ( \bar{\textbf{s} }^{(t)},\textbf{L}^{(t+1)} \right ) \right ]\\
&&& \stackrel{(a)}{\le} \left ( \gamma_l^2+L_l\gamma_l^2-\gamma_l  \right ) \Delta^{(t)}_{\textbf{l} } + \left ( \frac{\gamma_n}{2}+4L_s\gamma_s^2-\frac{\gamma_s}{2} \right ) \Delta^{(t)}_{\bar{\textbf{s} } }\\
&&& \quad +(1+L_l)\gamma_l^2\sigma_l^2+4L_s \gamma_s^2 (\sigma_s^2+ \sigma_g^2)\\
&&& \quad +\left ( \frac{L_{ls}^2}{2}+\frac{\gamma_sL_{sc}^2}{2}+4L_sL_{sc}^2\gamma_s^2 \right ) \frac{1}{N}\sum_{i=1}^{N}\mathbb{E}\left \|\textbf{y} ^{(t)}_{i}-\bar{\textbf{s} }^{(t)}  \right \| _2^2\\
&&& \quad +\left ( \gamma_n + 2L_s\gamma_s^2  \right )\frac{d_s S^2}{b^2},
\end{aligned}
\end{equation}

\noindent
where (a) is following Lemmas 4 and 5. Rearranging Eq.(\ref{theorem2eq1}), we have:
\begin{equation}\label{theorem2eq2}
\begin{aligned}
&&&\left (\gamma_l- \gamma_l^2-L_l\gamma_l^2  \right ) \Delta^{(t)}_{\textbf{l} } + \left (\frac{\gamma_s}{2}- \frac{\gamma_n}{2}-4L_s\gamma_s^2 \right ) \Delta^{(t)}_{\bar{\textbf{s} } }\\
&&& \le \mathbb{E} \left [ F\left ( \bar{\textbf{s} }^{(t)},\textbf{L}^{(t)} \right ) -F\left ( \bar{\textbf{s} }^{(t+1)},\textbf{L}^{(t+1)} \right )\right ]\\
&&& \quad +(1+L_l)\gamma_l^2\sigma_l^2+4L_s \gamma_s^2 (\sigma_s^2+ \sigma_g^2)\\
&&& \quad +\left ( \frac{L_{ls}^2}{2}+\frac{\gamma_sL_{sc}^2}{2}+4L_sL_{sc}^2\gamma_s^2 \right ) \frac{1}{N}\sum_{i=1}^{N}\mathbb{E}\left \|\textbf{y} ^{(t)}_{i}-\bar{\textbf{s} }^{(t)}  \right \| _2^2\\
&&& \quad +\left ( \gamma_n + 2L_s\gamma_s^2  \right )\frac{d_s S^2}{b^2}.
\end{aligned}
\end{equation}

\noindent
Summing up Eq.(\ref{theorem2eq2}) over $t\in \left \{ 0, 1,..., T-1 \right \}$, we have:
\begin{equation}\label{theorem2eq3}
\begin{aligned}
&&&\sum_{t=0}^{T-1}\left[\left (\gamma_l- \gamma_l^2-L_l\gamma_l^2  \right )\Delta^{(t)}_{\textbf{l} } + \left (\frac{\gamma_s}{2}- \frac{\gamma_n}{2}-4L_s\gamma_s^2 \right ) \Delta^{(t)}_{\bar{\textbf{s} } }\right]\\
&&& \le \mathbb{E} \left [ F\left ( \bar{\textbf{s} }^{(0)},\textbf{L}^{(0)} \right ) -F\left ( \bar{\textbf{s} }^{(T)},\textbf{L}^{(T)} \right )\right ]\\
&&& \quad +(1+L_l)\gamma_l^2\sigma_l^2T+4L_s \gamma_s^2 T(\sigma_s^2+ \sigma_g^2)\\
&&& \quad +\left ( \frac{L_{ls}^2}{2}+\frac{\gamma_sL_{sc}^2}{2} \right ) \frac{1}{N}\sum_{t=0}^{T-1}\sum_{i=1}^{N}\mathbb{E}\left \|\textbf{y} ^{(t)}_{i}-\bar{\textbf{s} }^{(t)}  \right \| _2^2\\
&&& \quad +\frac{4L_sL_{sc}^2\gamma_s^2}{N}\sum_{t=0}^{T-1}\sum_{i=1}^{N}\mathbb{E}\left \|\textbf{y} ^{(t)}_{i}-\bar{\textbf{s} }^{(t)}  \right \| _2^2\\
&&& \quad +\left ( \gamma_n + 2L_s\gamma_s^2  \right )\frac{d_s S^2T}{b^2}.
\end{aligned}
\end{equation}

\noindent
Defining $A:=\left ( \frac{L_{ls}^2}{2}+\frac{\gamma_sL_{sc}^2}{2}+4L_sL_{sc}^2\gamma_s^2 \right )\frac{30C^2}{(1-q)^2}$, by Lemma 6 and $\gamma_s <\min\left \{  \frac{1-q}{2\sqrt{15} CL_{sc}},\frac{1}{16L_s+8A}  \right \} $, we have:
\begin{equation}\label{theorem2eq4}
\begin{aligned}
&&&\left (\gamma_l- \gamma_l^2-L_l\gamma_l^2  \right ) \sum_{t=0}^{T-1}\Delta^{(t)}_{\textbf{l} } \\
&&& \quad + \left (\frac{\gamma_s}{2}- \frac{\gamma_n}{2}-4L_s\gamma_s^2-2A\gamma_s^2 \right ) \sum_{t=0}^{T-1}\Delta^{(t)}_{\bar{\textbf{s} } }\\
&&& \le \mathbb{E} \left [ F\left ( \bar{\textbf{s} }^{(0)},\textbf{L}^{(0)} \right ) -F\left ( \bar{\textbf{s} }^{(T)},\textbf{L}^{(T)} \right )\right ]+\frac{A}{3N}\sum_{i=1}^N\left\|\textbf{s}^{(0)}_i\right\|_2^2\\
&&& \quad +T\left[(1+L_l)\gamma_l^2\sigma_l^2+\left ( 4L_s+A \right ) \gamma_s^2 (\sigma_s^2+\sigma_g^2)\right]\\
&&& \quad +\left ( \gamma_n + 2L_s\gamma_s^2+2A\gamma_n^2  \right )\frac{d_s S^2T}{b^2}.
\end{aligned}
\end{equation}

\noindent
Since the upper bounds on the step sizes ensure that the left side of Eq.(\ref{theorem2eq4}) is positive, we consider $\gamma_l=O\left ( \frac{1}{\sqrt{T} }  \right )$, $\gamma_s=O\left ( \frac{1}{\sqrt{T} }  \right )$, and $\gamma_n=O\left ( \frac{1}{T}  \right )$, divide both sides of Eq.(\ref{theorem2eq4}) by $\sqrt{T}$. With sufficiently large $T$, we have the conclusion.
\begin{equation}\label{theorem2eq5}
\begin{aligned}
&&& O\left ( \frac{1}{T}\sum_{t=0}^{T-1}\Delta^{(t)}_{\textbf{l} }\right )+O\left ( \frac{1}{T} \sum_{t=0}^{T-1}\Delta^{(t)}_{\bar{\textbf{s} } }\right )\\ 
&&& \le O\left ( \frac{1}{\sqrt{T}} \right ) +O\left ( \frac{d_s S^2}{b^2\sqrt{T}} \right ) \\
&&&\quad +O\left ( \frac{\sigma_l^2}{\sqrt{T}} \right ) +O\left ( \frac{\sigma_s^2}{\sqrt{T}} \right ) +O\left ( \frac{\sigma_g^2}{\sqrt{T}} \right ) .
\end{aligned}
\end{equation}
\end{proof}

\section{Additional Experiments}
\subsection{Optimization Comparison of Real vs. Estimated Sensitivity in PartPSP}
\begin{table}[htbp]
\centering
\makeatletter
\makeatother
\caption{Performance Comparison: PartPSP-Real vs PartPSP-Esti}

\begin{tabular}{c|c|cc|cc}  
\toprule
\multicolumn{2}{c|}{Model/Dataset} & \multicolumn{2}{c|}{MLP/MNIST} & \multicolumn{2}{c}{ResNet/FMNIST} \\  
\cmidrule(lr){1-2} \cmidrule(lr){3-4} \cmidrule(lr){5-6}
\multicolumn{2}{c|}{Topology} & \multicolumn{1}{c}{2-Out} & \multicolumn{1}{c|}{EXP} & \multicolumn{1}{c}{2-Out} & \multicolumn{1}{c}{EXP} \\
\midrule
\multirow{2}{*}{PartPSP-Real} & 1-shared & 88.50 & 89.50 & 82.37 & 82.40 \\
& 2-shared & 60.53 & 61.14 & 80.83 & 81.21 \\
\midrule 
\multirow{2}{*}{PartPSP-Esti} & 1-shared & 85.42 & 84.88 & 82.28 & 81.56 \\
& 2-shared & 52.09 & 53.99 & 80.52 & 81.10 \\
\bottomrule
\end{tabular}
\label{RealVSEsti}
\end{table}

PartPSP uses the estimated sensitivity $S^{(t)}$ to generate privacy-preserving noise for differential privacy. Beyond validating the accuracy of $S^{(t)}$, an additional concern arises: to ensure rigorous privacy guarantees, $S^{(t)}$ must be conservatively overestimated. However, this leads to excessive noise injection, which may degrade optimization performance. A natural question is whether this performance cost is acceptable in practice. Under the same experimental setup as in Section V-B, we conducted experiments to address this question, and the results are reported in Table~\ref{RealVSEsti}. PartPSP-Real represents PartPSP using the real sensitivity for privacy, while PartPSP-Esti represents PartPSP using the estimated sensitivity. Table \ref{RealVSEsti} lists the models' test accuracy optimized by these two algorithms under different datasets, topologies, and partial communication strategies. The hyperparameter settings and detailed results are available in our open-source program.

As we can see, in the four MLP experiments, using the larger estimated sensitivity brings an average 7.76\% accuracy decrease to the models optimized by PartPSP-Esti compared with those optimized by PartPSP-Real. In the four ResNet-18 experiments, the models optimized by PartPSP-Esti have an average 0.41\% accuracy decrease compared with models optimized by PartPSP-Real. Considering all of the experiments, the estimated sensitivity brings an average 3.93\% degradation to PartPSP-Esti compared with PartPSP-Real. According to the results in Table \ref{RealVSEsti}, we believe that the optimization performance cost by PartPSP in using estimated sensitivity to strictly protect the node's data privacy is acceptable.

\subsection{Time Cost of PartPSP}

Although the two key techniques, sensitivity estimation and partial communication, have significantly reduced the time cost of PartPSP, their executions still have time overhead. Theoretically, the sensitivity estimation in Remark 1 introduces $O(N)$ communication and $O(d_s)$ computation, while partial communication reduces the communication complexity by $O((d-d_s)N)$. Therefore, compared to the original Push-SUM-based algorithm (SGP), PartPSP incurs an additional $O(d_s)$ computational and $O(N-(d-d_s)N)$ communication complexity. To investigate the practical impact of these factors on PartPSP's time cost, we conduct experiments and present the results in Table~\ref{TimeCostTable}.


Table~\ref{TimeCostTable} presents the time costs of three algorithms: the non-private Push-SUM SGD (SGP), the algorithm with sensitivity estimation but without partial communication (SGPDP), and the algorithm that incorporates both sensitivity estimation and partial communication (PartPSP-1). The privacy budget hyperparameter $b$ for SGPDP and PartPSP-1 is set to 3, while SGP is a special case of SGPDP without differential privacy. We also list the accuracy of these algorithms under the same experimental settings in Table~\ref{ACCLOSS} for comparison. All experiments are conducted on a platform equipped with an AMD Ryzen 9 7900X CPU and an NVIDIA GeForce RTX 4090 GPU. The communication bandwidth for each node in the network is set to 1 $Gbps$.

%

\begin{table*}[htbp]
\centering
\makeatletter
\makeatother
\caption{The Time Cost Experiments of PartPSP}
\begin{tabular}{c|c|cccc|cccc|cccc}
\toprule
\multicolumn{2}{c|}{Model/Data} & \multicolumn{4}{c|}{MLP/MNIST} & \multicolumn{4}{c|}{ResNet-18/FMNIST} & \multicolumn{4}{c}{ViT/Cifar10} \\
\cmidrule(lr){1-2} \cmidrule(lr){3-6} \cmidrule(lr){7-10} \cmidrule(lr){11-14}
\multicolumn{2}{c|}{Topology} & EXP & 10D4 & 10D6 & 10D8 & EXP & 10D4 & 10D6 & 10D8 & EXP & 10D4 & 10D6 & 10D8 \\
\midrule
\multirow{2}{*}{\centering SGP} & Time Cost (s) & \textbf{68.33} & \textbf{68.85} & \textbf{70.12} & \textbf{70.89} & 874.5 & 1260 & 1647 & 2034 & 226.3 & 253.5 & 281.7 & 318.3 \\
                        & ACC (\%)      & \textbf{81.55} & \textbf{81.64} & \textbf{81.62} & 81.62 & \textbf{90.18} & \textbf{90.31} & \textbf{89.98} & \textbf{90.16} & \textbf{78.65} & \textbf{78.47} & \textbf{78.48} & \textbf{78.46} \\
\midrule 
\multirow{2}{*}{\centering SGPDP} & Time Cost (s) & 78.39 & 78.93 & 80.92 & 81.92 & 1120 & 1504 & 1888 & 2275 & 275.9 & 293.3 & 322.0 & 364.7 \\
                          & ACC (\%)      & 29.78 & 26.94 & 29.34 & 45.63 & 71.65 & 72.12 & 71.01 & 72.83 & 19.43 & 29.55 & 51.64 & 68.96 \\
\midrule 
\multirow{2}{*}{\centering PartPSP-1} & Time Cost (s) & 74.64 & 74.36 & 75.68 & 76.22 & \textbf{145.7} & \textbf{147.6} & \textbf{154.4} & \textbf{159.4} & \textbf{138.1} & \textbf{137.9} & \textbf{138.1} & \textbf{140.2} \\
                              & ACC (\%)      & 48.08 & 63.51 & 80.12 & \textbf{88.87} & 82.73 & 81.86 & 82.06 & 82.59 & 71.78 & 76.24 & 76.72 & 76.98 \\
\bottomrule
\end{tabular}
\label{TimeCostTable}
\end{table*}

Comparing SGP and SGPDP, their time costs differ only in the sensitivity estimation and the noise addition introduced by differential privacy. The experimental results in Table 3 reveal that the introduced differential privacy in SGPDP not only increases the time cost of the algorithm, but also significantly degrades its optimization performance. Specifically, SGPDP's time cost is the highest in all experiments, with an average 16.58\% relative increase compared with SGP. Meanwhile, due to added noise, the test accuracy of models optimized by SGPDP decreases by an average of 41.17\% compared to those optimized by SGP. Thus, introducing differential privacy significantly degrades the original Push-SUM-based algorithm's performance.

Comparing SGP and PartPSP-1, the time cost of PartPSP-1 increases due to the introduced differential privacy but decreases with partial communication. The key to this trade-off is the number of shared parameters. Table \ref{TimeCostTable} shows that for the MLP model (fewer parameters), PartPSP's time cost is 8.16\% higher on average than SGP. In this case, the additional overhead introduced by differential privacy exceeds the time savings from partial communication. For larger models such as ResNet-18 and ViT, PartPSP reduces the average time cost by 83.15\% compared to SGP. In addition, despite added noise for differential privacy, the average test accuracy of models optimized by PartPSP-1 decreases by only 8.94\% relative to SGP, while that of SGPDP decreases by 41.17\%. This optimization performance improvement is attributed to partial communication, which effectively mitigates the performance degradation caused by differential privacy. Overall, benefiting from partial communication, PartPSP has a better capability in handling the trade-off between privacy protection and performance compared to traditional Push-SUM-based algorithms.
\end{document}